%% file: main.tex
\documentclass[11pt,a4paper]{article}
\usepackage{jheppub}
\usepackage{subfigure}
\usepackage{multirow}

\newcommand{\Et} {E_{\mathrm{T}}}
\newcommand{\AKT}{\rm{anti-}$k_{t}$}

\input{atlasphysics}

\usepackage{preprintcover}
\PreprintCoverPaperTitle{A search for flavour changing neutral currents in top-quark decays in $pp$ collision data collected with the ATLAS detector at $\sqrt{s}=7$~TeV}
\PreprintIdNumber{CERN-PH-EP-2012-139}
\PreprintCoverAbstract{\input{abstract}}
\PreprintJournalName{JHEP}

\title{\boldmath{A search for flavour changing neutral currents in top-quark decays in $pp$ collision data
collected with the ATLAS detector at $\sqrt{s}=7$~TeV}}

\author[a]{The ATLAS Collaboration}
\affiliation[a]{CERN, Geneva}
\emailAdd{atlas.publications@cern.ch}

\abstract{\input{abstract}}


\begin{document}
\maketitle

\input{intro.tex}
\input{data.tex}
\input{samples.tex}
\input{reco.tex}
\input{qzselection.tex}
\input{qzbackground.tex}
\input{systematics.tex}
\input{limits.tex}
\input{conclusions.tex}
\input{Acknowledgement-09Feb12-collisions.tex}

\bibliographystyle{JHEP}
\bibliography{fcnc_min}

\clearpage 
\input{atlas_authlist}

\end{document}

%% file: atlasphysics.tex
\chardef\letterchar=11
\chardef\otherchar=12
\chardef\eolinechar=5
\newcount\hrs\newcount\minu\newcount\temptime
\def\hm{\hrs=\time \divide\hrs by 60 \minu=\time\temptime=\hrs
\multiply\temptime by 60%
\advance\minu by -\temptime
\ifnum\minu<10 \let\zerofill=0\else \let\zerofill=\relax\fi
 \the\hrs:\zerofill\the\minu}
\def\lapprox{\ensuremath{\sim\kern-1em\raise 0.65ex\hbox{$<$}}}
\def\rapprox{\ensuremath{\sim\kern-1em\raise 0.65ex\hbox{$>$}}}

\def\antibar#1{\ensuremath{#1\bar{#1}}}

\def\ttbar{\antibar{t}}

\def\chic{\ensuremath{\raise.4ex\hbox{$\chi$}_{{c}}}} 
\def\chib{\ensuremath{\raise.4ex\hbox{$\chi$}_{{b}}}} 

\let\psii=\psi  
\def\psi{\ensuremath{\psii}}%

\def\ggino{\ensuremath{\mathchoice%
      {\displaystyle\raise.4ex\hbox{$\displaystyle\tilde\chi$}}%
         {\textstyle\raise.4ex\hbox{$\textstyle\tilde\chi$}}%
       {\scriptstyle\raise.3ex\hbox{$\scriptstyle\tilde\chi$}}%
 {\scriptscriptstyle\raise.3ex\hbox{$\scriptscriptstyle\tilde\chi$}}}}

\def\chinop{\ensuremath{\mathchoice%
      {\displaystyle\raise.4ex\hbox{$\displaystyle\tilde\chi^+$}}%
         {\textstyle\raise.4ex\hbox{$\textstyle\tilde\chi^+$}}%
       {\scriptstyle\raise.3ex\hbox{$\scriptstyle\tilde\chi^+$}}%
 {\scriptscriptstyle\raise.3ex\hbox{$\scriptscriptstyle\tilde\chi^+$}}}}
\def\chinom{\ensuremath{\mathchoice%
      {\displaystyle\raise.4ex\hbox{$\displaystyle\tilde\chi^-$}}%
         {\textstyle\raise.4ex\hbox{$\textstyle\tilde\chi^-$}}%
       {\scriptstyle\raise.3ex\hbox{$\scriptstyle\tilde\chi^-$}}%
 {\scriptscriptstyle\raise.3ex\hbox{$\scriptscriptstyle\tilde\chi^-$}}}}
\def\chinopm{\ensuremath{\mathchoice%
      {\displaystyle\raise.4ex\hbox{$\displaystyle\tilde\chi^\pm$}}%
         {\textstyle\raise.4ex\hbox{$\textstyle\tilde\chi^\pm$}}%
       {\scriptstyle\raise.3ex\hbox{$\scriptstyle\tilde\chi^\pm$}}%
 {\scriptscriptstyle\raise.3ex\hbox{$\scriptscriptstyle\tilde\chi^\pm$}}}}
\def\chinomp{\ensuremath{\mathchoice%
      {\displaystyle\raise.4ex\hbox{$\displaystyle\tilde\chi^\mp$}}%
         {\textstyle\raise.4ex\hbox{$\textstyle\tilde\chi^\mp$}}%
       {\scriptstyle\raise.3ex\hbox{$\scriptstyle\tilde\chi^\mp$}}%
 {\scriptscriptstyle\raise.3ex\hbox{$\scriptscriptstyle\tilde\chi^\mp$}}}}

\def\chinoonep{\ensuremath{\mathchoice%
      {\displaystyle\raise.4ex\hbox{$\displaystyle\tilde\chi^+_1$}}%
         {\textstyle\raise.4ex\hbox{$\textstyle\tilde\chi^+_1$}}%
       {\scriptstyle\raise.3ex\hbox{$\scriptstyle\tilde\chi^+_1$}}%
 {\scriptscriptstyle\raise.3ex\hbox{$\scriptscriptstyle\tilde\chi^+_1$}}}}
\def\chinoonem{\ensuremath{\mathchoice%
      {\displaystyle\raise.4ex\hbox{$\displaystyle\tilde\chi^-_1$}}%
         {\textstyle\raise.4ex\hbox{$\textstyle\tilde\chi^-_1$}}%
       {\scriptstyle\raise.3ex\hbox{$\scriptstyle\tilde\chi^-_1$}}%
 {\scriptscriptstyle\raise.3ex\hbox{$\scriptscriptstyle\tilde\chi^-_1$}}}}
\def\chinoonepm{\ensuremath{\mathchoice%
      {\displaystyle\raise.4ex\hbox{$\displaystyle\tilde\chi^\pm_1$}}%
         {\textstyle\raise.4ex\hbox{$\textstyle\tilde\chi^\pm_1$}}%
       {\scriptstyle\raise.3ex\hbox{$\scriptstyle\tilde\chi^\pm_1$}}%
 {\scriptscriptstyle\raise.3ex\hbox{$\scriptscriptstyle\tilde\chi^\pm_1$}}}}

\def\chinotwop{\ensuremath{\mathchoice%
      {\displaystyle\raise.4ex\hbox{$\displaystyle\tilde\chi^+_2$}}%
         {\textstyle\raise.4ex\hbox{$\textstyle\tilde\chi^+_2$}}%
       {\scriptstyle\raise.3ex\hbox{$\scriptstyle\tilde\chi^+_2$}}%
 {\scriptscriptstyle\raise.3ex\hbox{$\scriptscriptstyle\tilde\chi^+_2$}}}}
\def\chinotwom{\ensuremath{\mathchoice%
      {\displaystyle\raise.4ex\hbox{$\displaystyle\tilde\chi^-_2$}}%
         {\textstyle\raise.4ex\hbox{$\textstyle\tilde\chi^-_2$}}%
       {\scriptstyle\raise.3ex\hbox{$\scriptstyle\tilde\chi^-_2$}}%
 {\scriptscriptstyle\raise.3ex\hbox{$\scriptscriptstyle\tilde\chi^-_2$}}}}
\def\chinotwopm{\ensuremath{\mathchoice%
      {\displaystyle\raise.4ex\hbox{$\displaystyle\tilde\chi^\pm_2$}}%
         {\textstyle\raise.4ex\hbox{$\textstyle\tilde\chi^\pm_2$}}%
       {\scriptstyle\raise.3ex\hbox{$\scriptstyle\tilde\chi^\pm_2$}}%
 {\scriptscriptstyle\raise.3ex\hbox{$\scriptscriptstyle\tilde\chi^\pm_2$}}}}

\def\nino{\ensuremath{\mathchoice%
      {\displaystyle\raise.4ex\hbox{$\displaystyle\tilde\chi^0$}}%
         {\textstyle\raise.4ex\hbox{$\textstyle\tilde\chi^0$}}%
       {\scriptstyle\raise.3ex\hbox{$\scriptstyle\tilde\chi^0$}}%
 {\scriptscriptstyle\raise.3ex\hbox{$\scriptscriptstyle\tilde\chi^0$}}}}

\def\ninoone{\ensuremath{\mathchoice%
      {\displaystyle\raise.4ex\hbox{$\displaystyle\tilde\chi^0_1$}}%
         {\textstyle\raise.4ex\hbox{$\textstyle\tilde\chi^0_1$}}%
       {\scriptstyle\raise.3ex\hbox{$\scriptstyle\tilde\chi^0_1$}}%
 {\scriptscriptstyle\raise.3ex\hbox{$\scriptscriptstyle\tilde\chi^0_1$}}}}
\def\ninotwo{\ensuremath{\mathchoice%
      {\displaystyle\raise.4ex\hbox{$\displaystyle\tilde\chi^0_2$}}%
         {\textstyle\raise.4ex\hbox{$\textstyle\tilde\chi^0_2$}}%
       {\scriptstyle\raise.3ex\hbox{$\scriptstyle\tilde\chi^0_2$}}%
 {\scriptscriptstyle\raise.3ex\hbox{$\scriptscriptstyle\tilde\chi^0_2$}}}}
\def\ninothree{\ensuremath{\mathchoice%
      {\displaystyle\raise.4ex\hbox{$\displaystyle\tilde\chi^0_3$}}%
         {\textstyle\raise.4ex\hbox{$\textstyle\tilde\chi^0_3$}}%
       {\scriptstyle\raise.3ex\hbox{$\scriptstyle\tilde\chi^0_3$}}%
 {\scriptscriptstyle\raise.3ex\hbox{$\scriptscriptstyle\tilde\chi^0_3$}}}}
\def\ninofour{\ensuremath{\mathchoice%
      {\displaystyle\raise.4ex\hbox{$\displaystyle\tilde\chi^0_4$}}%
         {\textstyle\raise.4ex\hbox{$\textstyle\tilde\chi^0_4$}}%
       {\scriptstyle\raise.3ex\hbox{$\scriptstyle\tilde\chi^0_4$}}%
 {\scriptscriptstyle\raise.3ex\hbox{$\scriptscriptstyle\tilde\chi^0_4$}}}}

\let\pii=\pi
\def\pi{\ensuremath{\pii}}%
\let\etaa=\eta
\def\eta{\ensuremath{\etaa}}%
\def\pt{\ensuremath{p_{\mathrm{T}}}} 
\def\pT{\ensuremath{p_{\mathrm{T}}}} 
\def\ET{\ensuremath{E_{\mathrm{T}}}} 

\def\MET{\ensuremath{E_{\mathrm{T}}^{\mathrm{miss}}}} 
\def\met{\ensuremath{E_{\mathrm{T}}^{\mathrm{miss}}}} 

\def\TeV{\ifmmode {\mathrm{\ Te\kern -0.1em V}}\else
                   \textrm{Te\kern -0.1em V}\fi}%
\def\GeV{\ifmmode {\mathrm{\ Ge\kern -0.1em V}}\else
                   \textrm{Ge\kern -0.1em V}\fi}%
\def\MeV{\ifmmode {\mathrm{\ Me\kern -0.1em V}}\else
                   \textrm{Me\kern -0.1em V}\fi}%
\def\keV{\ifmmode {\mathrm{\ ke\kern -0.1em V}}\else
                   \textrm{ke\kern -0.1em V}\fi}%
\def\eV{\ifmmode  {\mathrm{\ e\kern -0.1em V}}\else
                   \textrm{e\kern -0.1em V}\fi}%
\let\tev=\TeV
\let\gev=\GeV

\def\TeVc{\ifmmode {\mathrm{\ Te\kern -0.1em V}/c}\else
                   {\textrm{Te\kern -0.1em V}/$c$}\fi}%
\def\GeVc{\ifmmode {\mathrm{\ Ge\kern -0.1em V}/c}\else
                   {\textrm{Ge\kern -0.1em V}/$c$}\fi}%
\def\MeVc{\ifmmode {\mathrm{\ Me\kern -0.1em V}/c}\else
                   {\textrm{Me\kern -0.1em V}/$c$}\fi}%
\def\keVc{\ifmmode {\mathrm{\ ke\kern -0.1em V}/c}\else
                   {\textrm{ke\kern -0.1em V}/$c$}\fi}%
\def\eVc{\ifmmode  {\mathrm{\ e\kern -0.1em V}/c}\else
                   {\textrm{e\kern -0.1em V}/$c$}\fi}%

\def\TeVcc{\ifmmode {\mathrm{\ Te\kern -0.1em V}/c^2}\else
                   {\textrm{Te\kern -0.1em V}/$c^2$}\fi}%
\def\GeVcc{\ifmmode {\mathrm{\ Ge\kern -0.1em V}/c^2}\else
                   {\textrm{Ge\kern -0.1em V}/$c^2$}\fi}%
\def\MeVcc{\ifmmode {\mathrm{\ Me\kern -0.1em V}/c^2}\else
                   {\textrm{Me\kern -0.1em V}/$c^2$}\fi}%
\def\keVcc{\ifmmode {\mathrm{\ ke\kern -0.1em V}/c^2}\else
                   {\textrm{ke\kern -0.1em V}/$c^2$}\fi}%
\def\eVcc{\ifmmode  {\mathrm{\ e\kern -0.1em V}/c^2}\else
                   {\textrm{e\kern -0.1em V}/$c^2$}\fi}%

\def\cm{\ifmmode  {\mathrm{\ cm}}\else
                   \textrm{~cm}\fi}%
\newbox\boxsqbox
\newdimen\boxsize\boxsize=1.2ex%
\def\boxop{%
\setbox\boxsqbox=\vbox{\hrule depth0.8pt width0.8\boxsize height0pt%
                       \kern0.8\boxsize
                       \hrule height0.8pt width0.8\boxsize depth0pt}%
           \hbox{%
           \vrule height1.0\boxsize width0.8pt depth0pt%
           \copy\boxsqbox
           \vrule height1.0\boxsize width0.8pt depth0pt\kern1.5pt}}%
\def\pmb#1{\setbox0=\hbox{$#1$}
  \kern-.025em\copy0\kern-1.0\wd0%
  \kern.05em\copy0\kern-1.0\wd0%
  \kern-.025em\raise.0433em\box0}%
\newdimen\dkwidth
\def\dk{%
   \dkwidth=\baselineskip
   {\def\to{\rightarrow}
   \kern 3pt%
   \hbox{%
      \raise 3pt%
      \hbox{%
         \vrule height 0.8\dkwidth width 0.7pt depth0pt%
      }%
      \kern-0.4pt%
      \hbox to 1.5\dkwidth{%
         \rightarrowfill
      }%
   \kern0.6em%
   }}%
}%
\def\eqalign#1{%
   \,
   \vcenter{%
      \openup\jot\m@th
      \ialign{%
         \strut\hfil$\displaystyle{##}$&&$%
         \displaystyle{{}##}$\hfil\crcr#1\crcr%
      }%
   }%
   \,
}%

%% file: abstract.tex
A search for flavour changing neutral current~(FCNC) processes in
top-quark decays by the ATLAS Collaboration is presented. Data
collected from $pp$ collisions at the LHC at a centre-of-mass energy of
$\sqrt{s}=7$~\tev\ during 2011, corresponding to an integrated
luminosity of 2.1~fb$^{-1}$, were used. A search was performed for
top-quark pair-production events, with one top quark decaying
through the $t\to Zq$ FCNC ($q=u,c$) channel, and the other through
the Standard Model dominant mode $t\to Wb$. Only the decays of the
$Z$ boson to charged leptons and leptonic $W$-boson decays were
considered as signal. Consequently, the final-state topology is
characterised by the presence of three isolated charged leptons, at
least two jets and missing transverse momentum from the undetected
neutrino. No evidence for an FCNC signal was found. An upper limit
on the $t\to Zq$ branching ratio of BR$(t\to Zq)<0.73$\% is set at
the 95\% confidence level.

%% file: intro.tex
\section{Introduction\label{sec:intro}}

The top quark is the heaviest known elementary particle, with a mass
of $m_t=173.2\pm0.9$~\gev\ ~\cite{TevEWKWG:2011wr}. The very large
mass may provide a window onto physics beyond the Standard Model
(SM). Deviations from SM predictions of the production and decay
properties of the top quark provide model-independent tests for
physics beyond the SM.
 According to the SM, the top quark decays nearly 100\% of the time to a $W$ boson and a $b$ quark.
  Flavour changing neutral current (FCNC) decays are highly
  suppressed in the SM by the GIM mechanism~\cite{Glashow:1970gm} with a branching ratio (BR) of the order of $10^{-14}$.

Several SM extensions predict a higher BR for top-quark FCNC decays.
Examples of such extensions are the quark-singlet
model~\cite{AguilarSaavedra:2002ns, delAguila:1998tp,
 AguilarSaavedra:2002kr}, the two-Higgs doublet model with or without
flavour-conservation~\cite{cheng:1987rs, Grzadkowski:1990sm,
Luke:1993cy, Atwood:1995ud, Atwood:1996vj, Bejar:2000ub}, the
minimal supersymmetric model~\cite{Li:1993mg, deDivitiis:1997sh,
Lopez:1997xv, Guasch:1999jp, Delepine:2004hr, Liu:2004qw,
Cao:2007dk}, supersymmetry (SUSY) with $R$-parity
violation~\cite{Yang:1997dk}, the topcolour-assisted technicolour
model~\cite{Lu:2003yr} or models with warped extra
dimensions~\cite{Agashe:2004cp,Agashe:2006wa}. The top-quark FCNC
decay BR in these models is typically many orders of magnitude
larger than the SM BR, and can be as high as $\sim 2\times 10^{-4}$
in certain $R$-parity violating SUSY models.

Existing experimental limits on top-quark FCNC decays come from
direct and indirect searches at the Tevatron
collider~\cite{Abe:1997fz,Abazov:2011qf}, and indirect searches at
the LEP~\cite{Heister:2002xv, Abdallah:2003wf, Abbiendi:2001wk,
Achard:2002vv, LEPExoticaWG200101,Beneke:2000hk} and
HERA~\cite{Abramowicz:2011tv,Aktas:2003yd} colliders, and at the
LHC~\cite{Aad2012351}. The best current direct search limits on the
top quark FCNC branching fraction are 3.2\% for both $t\rightarrow
q\gamma$~\cite{Abe:1997fz} and $t\rightarrow Zq$
($q=u,c$)~\cite{Abazov:2011qf}.

This article reports a search for FCNC top-quark decays in $\ttbar$
events. Events were  searched for in which either the top or antitop
quark has decayed into a $Z$ boson and a quark, $t\rightarrow Zq$,
while the remaining top or antitop quark decayed through the SM
$t\rightarrow Wb$ channel. Only leptonic decays of the $Z$ and $W$
bosons were considered, yielding a final-state topology
characterised by the presence of three isolated charged leptons, at
least two jets, and transverse momentum imbalance ($\MET$) from the
undetected neutrino arising from the $W$-boson decay.  Leptons are
either well-identified electron or muon candidates, selected using
the full detector or, to increase signal acceptance, isolated
tracks. Channels with $\tau$ leptons are not explicitly
reconstructed, but reconstructed electrons and muons can arise from
leptonic $\tau$ decays, and an isolated track can arise from
hadronic $\tau$ decay modes.

%% file: data.tex
\section{Detector and data samples\label{sec:data}}

The ATLAS detector~\cite{Aad:2008zzm} at the LHC covers nearly the
entire solid angle around the collision point. It consists of an
inner tracking detector comprising a silicon pixel detector, a
silicon microstrip detector (SCT), and a transition radiation
tracker. The inner detector covers the pseudorapidity\footnote{In
the right-handed ATLAS coordinate system, the
  pseudorapidity $\eta$ is defined as $\eta=-\ln[\tan(\theta/2)]$,
  where the polar angle $\theta$ is measured with respect to the LHC
  beamline. The azimuthal angle $\phi$ is measured with respect to the
  $x$-axis, which points towards the centre of the LHC ring.  The
  $y$-axis points upwards.  Transverse momentum and energy are defined as
  $\pT = p\sin\theta$ and $\Et = E\sin\theta$, respectively.}\ range $|\eta |<2.5$ and is surrounded by a thin superconducting
solenoid providing a 2~T axial magnetic field, and by lead/liquid-argon (LAr)
electromagnetic (EM) sampling calorimeters with high granularity. An iron/scintillator-tile calorimeter provides hadronic
energy measurements in the central pseudorapidity range ($|\eta|<
1.7$). The end-cap and forward regions are instrumented with LAr
calorimeters for both EM and hadronic energy measurements up to
$|\eta|< 4.9$. The calorimeter system is surrounded by a muon
spectrometer incorporating three superconducting toroid magnet
assemblies (one barrel and two end-caps), with bending power between
2.0 Tm and 7.5 Tm.

A three-level trigger system is used to collect data. The
first-level trigger is implemented in hardware and uses a subset of
the detector information to reduce the rate to at most 75\,kHz. This
is followed by two software-based trigger levels that together
reduce the event rate to $\sim 300$ Hz.  This analysis uses
inclusive single-muon and single-electron triggers with $\pT$
thresholds of 18 \gev\ for muons and 20 \gev\ or 22 \gev\ for
electrons, depending on the data taking period.

Proton-proton collision data taken at $\sqrt{s}=7$~TeV by ATLAS
between March and August 2011 are used.  The data sample corresponds
to a total integrated luminosity of 2.1~fb$^{-1}$ with an
uncertainty of 3.7\%~\cite{lumiPub, lumi}. The mean number of
interactions per bunch crossing was 6.2 for the full data sample.

%% file: samples.tex
\section{Monte Carlo simulation samples}
\label{sec:MC_Samples}
Monte Carlo (MC) samples were generated to model both FCNC signal
events and certain backgrounds.  Alternative MC samples were also
generated to evaluate various systematic uncertainties. All
generated events are propagated through a detailed {\tt GEANT4}
simulation~\cite{Agostinelli2003250, ATLAS_SIM} of the ATLAS
detector and are reconstructed with the same algorithms as the data.
The effect of additional $pp$ interactions in the same bunch
crossing as the events of interest was simulated by superimposing
additional simulated minimum-bias interactions. The effect from
events in neighbouring bunch crossings was also simulated.  The
simulated events were reweighted such that the average number of
extra interactions per crossing (pile-up) matched the data.
Data-to-MC scale factors were applied to the MC samples to account
for small differences in efficiencies between data and MC
simulation.
\subsection{Signal}

Monte Carlo simulation samples of top-quark pair production, with
one of the top quarks decaying through FCNC to $Zq$ while the other
decays according to the SM, were generated with {\tt
TopReX}~\cite{Slabospitsky:2002ag}.
 Only decays of the $W$ and
$Z$ bosons involving charged leptons were generated ($Z\to
ee,\mu\mu,\tau\tau$ and $W\to e\nu,\mu\nu,\tau\nu$). The
MRST2007 LO$^*$~\cite{Sherstnev:2007nd} parton distribution function (PDF)
set was used with the {\tt TopReX} generator. All signal events were
hadronized with {\tt PYTHIA}~6.421~\cite{Sjostrand:2006za}. The
masses of the top quark, $W$ boson and $Z$ boson were set to
172.5~\gev, 80.4~\gev\ and 91.2~\gev, respectively.

To study the effect of the uncertainty due to the top-quark mass,
samples with top-quark masses of 170~\gev\ and 175~\gev\ were also
generated. The uncertainty due to initial- and final-state radiation
(ISR/FSR) was evaluated using the {\tt AcerMC}
generator~\cite{Kersevan:2004yg} interfaced to {\tt PYTHIA}, and by
varying the parameters controlling ISR and FSR in a range consistent
with those used in the Perugia Hard/Soft tune
variations~\cite{Skands}.

\subsection{Background}
\label{sec:Bkg}

Several SM processes have final-state topologies similar to the
signal. These include events with three final-state charged leptons
(real leptons), as well as events in which at least one jet
(including jets with heavy-flavour decays) is misidentified as an
isolated
charged lepton (`fake leptons') and events with four leptons in which one is not reconstructed. 

Diboson events ($WW$, $WZ$ and $ZZ$) were produced using {\tt
ALPGEN}~2.13~\cite{alpgenRepeated}. Up to three additional partons
from the matrix element were simulated, and the
CTEQ6L1~\cite{cteq6l} PDF was used. The parton shower and the
underlying event were added using {\tt HERWIG} v6.510~\cite{herwig1, Corcella:2002jc} and the {\tt
JIMMY}~\cite{Butterworth:1996zw} underlying event model with the
AUET1 tune~\cite{PUB-2010-014} to the ATLAS data.  The {\tt ALPGEN} program with {\tt HERWIG} showering and the {\tt JIMMY} underlying event model was also used to generate $Z/\gamma +$jets.

The \ttbar\ and single-top events were generated using the {\tt
MC@NLO} generator
v3.41~\cite{mcatnlo1,Frixione:2003ei,Frixione:2005vw} with the
CTEQ6.6~\cite{Nadolsky:2008zw} PDFs~\cite{cteq66}. The parton shower and the underlying event
were added
using {\tt HERWIG} v6.510 and {\tt JIMMY} 
generators as described above. The \ttbar\ production cross section
was normalized to the approximate next-to-next-to-leading-order
(NNLO) prediction of 164.6~pb, obtained using the {\tt HATHOR} tool
\cite{Aliev:2010}. The cross sections for single-top production were
normalized to the approximate NNLO predictions of 64.6~pb
\cite{Kidonakis:2011wy}, 4.6~pb \cite{Kidonakis:2010tc} and 15.7~pb
\cite{Kidonakis:2010ux} for $t$-channel, $s$-channel and associated
$Wt$ production, respectively.

Events with \ttbar +$W$ and \ttbar +$Z$ production, including those with extra jets
in the final state, were generated using {\tt
MADGRAPH}~4.4.62~\cite{Alwall:2007st}.  Parton showering was added using {\tt PYTHIA}.

 All decay modes of the $W$ and $Z$ bosons to charged leptons
were considered in the generation and simulation of the background
samples used.

Backgrounds that include fake leptons were evaluated using a data-driven
approach described below.


%% file: reco.tex
\section{Object definition\label{sec:reco}}

The selection of leptons, jets, and $\MET$ was close to that
used for the ATLAS measurement of the \ttbar\ production cross
section in the dilepton channel~\cite{DILPaper}.  Leptons were
selected either using the full ATLAS detector, including the inner
detector, calorimeter and muon spectrometer (`identified leptons' or
`ID leptons'), or using only a high quality inner detector track
(`track leptons' or `TLs').  The inclusion of TLs increased the
acceptance for $W\rightarrow\tau\nu$ decays, and for electrons and
muons that fail the ID lepton selection criteria.  TLs were required
to be distinct from ID leptons, and at most one TL per event was
allowed. Signal candidates selected with three identified leptons
are referred to as `3ID' events, and those selected with two
identified leptons and a TL are referred to as `2ID+TL' events. The
2ID+TL events increased the signal acceptance by 22\% compared to a
3ID selection alone.

ID electron candidates were reconstructed from energy deposits
(clusters) in the EM calorimeter, which were then associated to
reconstructed tracks of charged particles in the inner detector.
Stringent quality requirements on the conditions of the EM
calorimeter at the time of data taking were applied to ensure a
well-measured reconstructed energy. A `tight'
selection~\cite{ElectronPerformance} using calorimeter, tracking and
combined variables, was employed to provide good separation between
the signal electrons and background.  Electron candidates were
additionally required to have $|\eta_{\mathrm{cl}}|<2.47$, excluding
electrons in the transition region between the barrel and endcap
calorimeters defined by $1.37<|\eta_{\mathrm{cl}}|<1.52$. The
variable $\eta_{\mathrm{cl}}$\ is the pseudorapidity of the energy
cluster associated with the candidate. 

ID muon candidate reconstruction began by searching for track
segments in layers of the muon chambers. These segments were
combined starting from the outermost layer, fitted to account for
material effects, and matched with tracks found in the inner
detector. The candidates were refitted using the complete track
information from both detector systems, and were required to satisfy
$|\eta|<2.5$.

Candidates for TL were defined by an inner-detector track and a
series of quality cuts optimized for high efficiency and a low rate
of misidentification. The track was required to have at least six
pixel and/or SCT hits and at least one hit in the innermost pixel
layer. The transverse distance of closest approach of the track to
the beamline, $d_0$, was required to satisfy $|d_0| < 0.2$~mm and
the uncertainty on the momentum measurement was required be less
than 20\%.

All leptons were required to be isolated and have high transverse
momentum, $\pt$, consistent with originating from $W$- or $Z$-boson
decay.  Because of the requirement of three leptons in this
analysis, lepton thresholds were reduced from those used in
Ref.~\cite{DILPaper}.  In 3ID events, the leading lepton was
required to have $\pt >25$~\gev, and the two sub-leading leptons
were required to have $\pt >20$~\gev.  In 2ID+TL events, the TL was
required to have $\pt>25$~\gev, and the two ID leptons in the event
were required to have $\pt >20$~\gev.  At least one ID lepton was
required to have fired the trigger.  To ensure good trigger
efficiency with the higher electron trigger thresholds,
reconstructed electrons that were associated with trigger objects
were required to have $\pt>25$~\gev.

Lepton isolation requirements reduce backgrounds from misidentified
jets and suppress the selection of leptons from heavy-flavour
decays. For ID electron candidates, $\ET$ deposited in the
calorimeter cells but not associated to the electron was summed in a
cone with radius\footnote{$\Delta R\equiv\sqrt{(\Delta\eta)^2 +
(\Delta\phi)^2}$} $\Delta R = 0.2$ around the electron and required
to be less than $3.5$ \gev. For ID muon candidates, the isolation
requirement was based on both calorimeter and track information. The
track isolation requirement was based on the sum of the track
transverse momenta, for tracks with $\pt >1$ \gev\ in a cone with
radius $\Delta R = 0.3$\ centred on the muon candidate, while the
calorimeter isolation requirement was based on the transverse energy
in the same cone. Both the track and calorimeter sums, excluding the
muon candidate, were required to be less than 4~\gev. Additionally,
ID muon candidates were required to have a distance $\Delta R >
0.4$\ from any jet with $\pt>20$~\gev, further suppressing muon
candidates from heavy-flavour decays.
 For TLs, the track was required to be isolated
from other nearby tracks following the track isolation definition
above, in this case using tracks with $\pt >0.5$ \gev. The summed
momentum cut was set to 2~\gev.  ID muon candidates arising from
cosmic rays were rejected by removing candidate pairs that were
back-to-back in the $r-\phi$ plane and with transverse impact
parameters relative to the beam axis $|d_0|~>~0.5~$mm.

Jets were reconstructed with the {\it \AKT} algorithm~\cite{antikt}
with a radius parameter $R = 0.4$, starting from energy clusters in
the calorimeter reconstructed using the scale established for
electromagnetic objects. These jets were then calibrated to the
hadronic energy scale using $\pt$- and $\eta$-dependent correction
factors~\cite{jetcor}. Jets were removed if they are within $\Delta
R<0.2$\ of a well-identified electron candidate, or within $\Delta R<0.4$\
of a TL. The jets used in the analysis were required to have
$\pt>25$~\gev\ and $|\eta|<2.5$.

To suppress backgrounds in which TLs are reconstructed from fake
leptons, a jet consistent with originating from a $b$ quark was
required in events with a TL. Jets were identified as $b$-quark
candidates (`$b$-tagged') by an algorithm that forms a likelihood
ratio of $b$- to light-quark jet hypotheses using several
kinematic variables~\cite{AdvBtag}. The cut on the combined
likelihood ratio was chosen such that a $b$-tagging efficiency of
$\approx 80\%$\ per $b$-jet in \ttbar~candidate events was achieved.

The $\MET$ vector was formed from the negative vector sum of the
transverse momenta of the reconstructed objects (electrons, muons,
jets)~\cite{METPaper}. The contribution from cells associated with
electron candidates was replaced by the calibrated transverse energy
of the candidate. The contribution from all ID muon candidates and
calorimeter clusters (including those not belonging to a
reconstructed object) was also included. TL candidates that arise
from muons and leave little energy in the calorimeter are not
properly included in the $\met$.  In such events the $\met$ often
points close to the TL direction.  In these events the $\met$ was
corrected with the $\pt$ of the TL if the TL and an
oppositely-charged ID muon were consistent with coming from a
$Z$-boson decay, and if the $\Delta\phi$ between the $\met$ and the
TL direction was less than 0.15 and there is no ID lepton within
$\Delta R$=0.05 of the TL (in which case the correction was already
included by the $\met$ algorithm). After all corrections, $\met >$
20 \gev\ was required.

%% file: qzselection.tex
\section{Event selection and reconstruction}
\label{s:event} The analysis required collision data selected by an
inclusive single-electron or single-muon trigger. To ensure that the
event was triggered by the lepton candidates used in the analysis,
one of the identified leptons and the triggered lepton were required
to match within $\Delta R < 0.15$.

Events were required to have a primary interaction vertex with at
least five tracks with $\pt>400\MeV$.  The event was discarded if
it had any jet with $\pt>20$ \gev\ that failed quality cuts designed to reject jets
arising from calorimeter noise or activity inconsistent with the
bunch-crossing time~\cite{jetcor}. If an electron candidate and a
muon candidate shared a track, the event was also discarded.

During part of the
data-taking period, corresponding to an integrated luminosity of
0.9~fb$^{-1}$, an electronics failure in a small $\eta - \phi$
region of the LAr EM calorimeter created a dead region.  For this
subset of the data, events in data and MC simulation containing
either an identified electron or a jet with $\pt >20$~\gev,
satisfying $-0.1 <~\eta ~< 1.5$ and $-0.9 <~\phi ~< -0.5$ were
rejected.

\begin{figure}
\begin{center}

\centering \subfigure[]{
\includegraphics[width=0.4\textwidth]{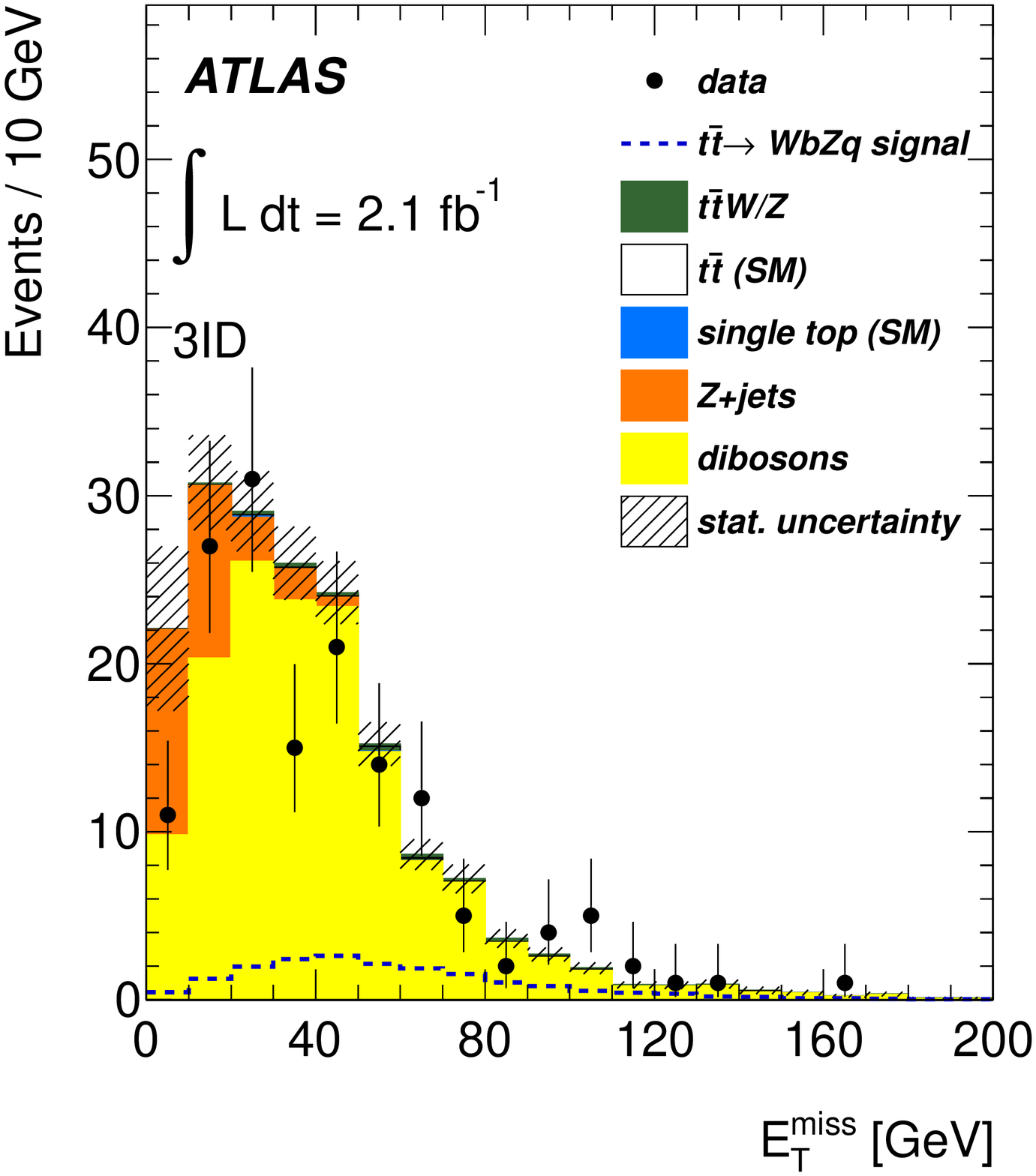}
}
\centering \subfigure[]{
\includegraphics[width=0.4\textwidth]{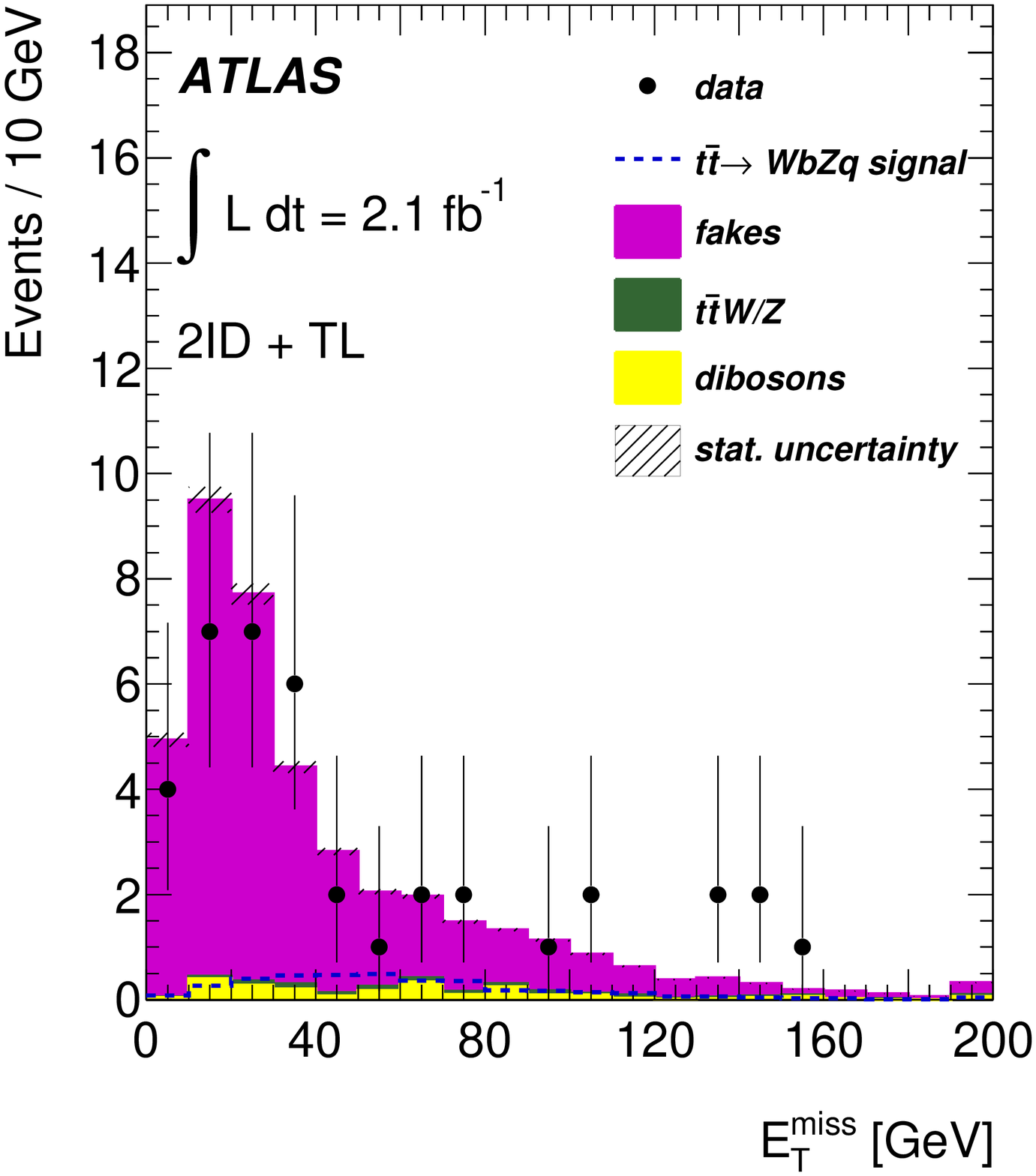}
} \caption{\label{fig:presel}$\met$ distributions before the final
selection for the (a) 3ID and (b) 2ID+TL analysis. For the 3ID case
these are events with three identified leptons with at least one
opposite-sign, same-flavour pair with an invariant mass consistent
with a $Z$-boson, but no jets or $\met$ requirement.  In the 2ID+TL
case, these are events with an opposite-sign pair and both the jet
and $\met$ requirements, but no $Z$-boson selection.  The
uncertainties shown are statistical only.}
\end{center}
\end{figure}

Events were selected as either 3ID or 2ID+TL candidates, each with
thresholds described in Section~\ref{sec:reco}.  In both cases, all
three lepton candidates were required to come from the same primary
interaction vertex.   In addition, signal candidates were required
to have at least two jets and $\met >$ 20 \gev. In events selected
with a TL, at least one jet was required to be $b$-tagged.
Figure~\ref{fig:presel} shows the $\met$ distribution for the 3ID
and 2ID+TL events prior to the final selection requirements.  For
the 3ID case these are events with three identified leptons with at
least one opposite-sign, same-flavour pair with an invariant mass
consistent with a $Z$-boson, but no jets or $\met$ requirement.  In
the 2ID+TL case, these are events with an opposite-sign pair and
both the jet and $\met$ requirements, but no $Z$-boson selection.

Selected events were required to be kinematically consistent with
$\ttbar\rightarrow WbZq$ through a $\chi^2$ minimized with respect
to jet and lepton assignments and the longitudinal momentum of the
neutrino, $p_z^\nu$. The $\chi^2$ was defined as follows

\begin{equation}
  \chi^2 =
  {\left(m^{\mathrm{reco}}_{j_a\ell_a\ell_b}  - m_t\right)^2\over \sigma_{t}^2} +
  {\left(m^{\mathrm{reco}}_{j_b\ell_c\nu}     - m_t\right)^2\over \sigma_{t}^2} +
  {\Big(m^{\mathrm{reco}}_{\ell_c\nu}        - m_W\Big)^2\over \sigma_{W}^2} +
  {\left(m^{\mathrm{reco}}_{\ell_a\ell_b}     - m_Z\right)^2\over \sigma_{Z}^2},
\label{qz3l:eq:chi2}
\end{equation}

\noindent where $j_{a,b}$ are the two highest-$\pt$ jets in the
event and $\ell_{a,b,c}$ are the three lepton candidates. The
constraints were defined as follows: $m_t=172.5$~\gev,
$m_W=80.4$~\gev, $m_Z=91.2$~\gev. The widths were derived from MC
studies and set to $\sigma_{t}=14$~\gev, $\sigma_{W}=10$~\gev\ and
$\sigma_{Z}=3$~\gev. The transverse momentum of the neutrino was set
equal to $\met$, and all jet and lepton assignments were tried,
subject to the requirement that the $Z$ candidate be built from
same-flavour opposite-charge leptons. Any opposite-charge ID
lepton--TL pair can be used as leptons from the $Z$-boson decay,
because the TL is assumed to be the same flavour as the ID lepton.
No $b$-jet identification was used in the reconstruction of the
event kinematics. For each assignment, the value of $p_z^\nu$ was
defined to be that which gave the minimum $\chi^2$.  From all
combinations, the one with the smallest $\chi^2$ was chosen along
with the corresponding $p_z^\nu$ value. Events were rejected unless
the reconstructed top-quark masses were within 40~\gev\ of $m_t$,
the reconstructed $W$-boson mass was within 30~\gev\ of $m_W$, and
the reconstructed $Z$-boson mass was within 15~\gev\ of $m_Z$. The
effect of these mass cuts on the fake-TL background expectation was
derived from simulation by measuring the fraction of simulated
2ID+TL background events with fake TLs that pass the $\chi^2$ mass
cuts.  This fraction is (31$\pm$10)\%.  Of the events that pass all
other event selection requirements, 38\% (29\%) of the 3ID (2ID+TL)
events pass the $\chi^2$ mass cuts. The efficiency for FCNC MC
events to pass the $\chi^2$-based mass cuts is $(79\pm 2)\%$
($(66\pm 2)$\%), while for background MC events it is only $(47\pm
7)\%$ ($(33\pm 10)$\%) for 3ID (2ID+TL) events.

The signal efficiency for $\ttbar\rightarrow WbZq$, after all
selection requirements, was determined using the {\tt TopReX} sample
described in Section~\ref{sec:MC_Samples} and is shown in
Table~\ref{tab:sel}.

%% file: qzbackground.tex
\section{Background evaluation}
Backgrounds to this search can be divided into two categories: those
with three real leptons and those with at least one fake lepton.
Backgrounds with three real leptons arise from diboson ($WZ$ and
$ZZ$) production with additional jets, and were evaluated using the
MC samples described in Section~\ref{sec:Bkg}. In the case of $WZ$
production, the required $\met$ comes from the neutrino from the
leptonic $W$-boson decay. Events from $ZZ$ decays can enter the
signal region in several ways; the dominant modes are four-lepton
decays with one lepton not reconstructed, giving apparent $\met$,
and $\tau^+\tau^-$ decays with one $\tau$ decaying to $e$ or $\mu$
and two neutrinos.

The background to 3ID candidate events, in which exactly one of the leptons
is a fake lepton, was evaluated using a combination of data and MC
samples.  The dominant contribution in this category comes from
$Z$+jets events, with a leptonic $Z$ decay, in which one of the jets
was misidentified as a third lepton.  To evaluate this background a
data-driven (DD) method was used. This method uses a control region (CR) in the
($\met$, $m_{\ell \ell}$) plane by selecting events with exactly two opposite-charge electrons or muons (no third ID lepton is allowed)
and $|91.2$~\gev$-m_{\ell\ell}^{\mathrm{reco}}|<15$~\gev\ in six
different $\met$ bins from 0 GeV to $\geq$50 GeV. The $Z+$jets
estimate in each $\met$ bin is then given by:
\begin{equation}
[N_{Z+\mathrm{jets}}^{\mathrm{Data}}]_{\mathrm{SR}}=
\left[\frac{N^{\mathrm{Data}}-N_{\mathrm{Other~backgrounds}}^{\mathrm{MC}}}{N_{Z+\mathrm{jets}}^{\mathrm{MC}}}\right]_{\mathrm{CR}}\cdot
\left[N_{Z+\mathrm{jets}}^{\mathrm{MC}}\right]_{\mathrm{SR}}. \label{qf:eq:DD}
\end{equation}
For each $\met$ bin considered, the corresponding
background-subtracted data/simulation ratio in the CR was applied to
the simulated $Z+$jets background in the signal regions (SR), in
order to evaluate the expected number of $Z+$jets events in the
data.  To enhance the statistical power of the MC sample, the lepton
selection was loosened compared to the SR lepton selection.  A
multiplicative factor of 0.063$\pm$0.013, corresponding to the MC
probability for events with loose leptons to pass the SR lepton
criteria, was applied to the final result.  The remaining
backgrounds with one fake lepton (dileptonic $t\bar t$, $Wt$-channel
single-top and $WW$ production) were evaluated using Monte Carlo
simulation samples, described in Section~\ref{sec:MC_Samples}, and
the loose lepton selection and multiplicative factor above.
Different DD methods and cross-checks for the one-fake-lepton
background were performed. These include the matrix
method~\cite{top2010}, relaxation of $\met$ or lepton quality
requirements, and MC simulation with fake-rate factors measured from
data. These alternative methods, although statistically limited,
agree with the reference DD+MC method used.

 A DD method was developed to evaluate the contribution to 3ID events
 from multi-jet, $W$+jets, single-top and $t\bar t$ single-lepton decay
events, in which two or three jets were reconstructed as
leptons~(2+3 fake leptons). Due to the requirement that two leptons
should have the same flavour and opposite charges, the yield from
these backgrounds can be extrapolated from the number of observed
data events with three leptons of any flavour ($e$ or $\mu$), but
with the same charge. Taking into account the possible charge and
flavour combinations, there are 36 combinations of three leptons,
in which two have the same flavour and opposite charges, and 16
combinations of three leptons with the same charge. The
extrapolation factor is thus $f=36/16=2.25$. No data event passed
the selection after requiring three leptons with the same charge.
The uncertainties in the DD backgrounds were determined using the
Feldman-Cousins upper interval for a 68\% C.L.~\cite{Feldman:1997qc}
with no observed events (with the uncertainties multiplied by 2.25
for the 2+3 fake leptons sample). Since no events were selected with
three leptons of the same charge, a multiplicative factor of
$0.071~\pm~0.018$, to account for the final requirements of at least
two jets with $\pt>$ 25 \gev\ and $\met >$20 \gev\, was evaluated
using MC samples and
 applied to the uncertainty estimate.

In 2ID+TL events the dominant background contribution comes from
events with a fake TL.  The background contribution from such events
was evaluated with the same technique used in Ref.~\cite{DILPaper}: the
probability of a jet being reconstructed as a track lepton was
determined from a $\gamma$+jets data sample selected with photon
triggers, and parameterized in a `fake matrix' as a function of jet
$\pT$ and the number of primary vertices in the event, $N_{\rm
vtx}$. The number of primary vertices was needed in the
parameterization because the fake probability is sensitive to
pile-up.  The fake matrix was applied to a `parent sample' selected
with all of the signal region requirements with the exception of the
three leptons. Instead, two ID leptons were required.  Fake
probabilities from the matrix were summed for each jet in the parent
sample, according to its $\pT$ and $N_{\rm vtx}$ for each event. The
resulting sum is the fake TL background contribution. Because of the
$b$-tag requirement in events with a TL, a $b$-tagged jet was
allowed to contribute to the sum  only if there was another
$b$-tagged jet in the event. This accounts for the fact that if a
jet produced a fake TL, the remaining jet would be removed by the
lepton--jet overlap removal described in Section~\ref{sec:reco}. For
the same reason, events with three or more jets were used to predict
the number of fake TLs in events with two or more jets. The signal
region required a $Z$-boson candidate, i.e. an opposite-charge, same-flavour lepton pair. Therefore the parent sample with two ID leptons
provides three different cases:
\begin{enumerate}
  \item Opposite-charge ID leptons
  \item Two positively-charged ID leptons
  \item Two negatively-charged ID leptons
\end{enumerate}

In case 1, the fake TL was allowed to have either charge.  In case 2
the fake TL was required to be negatively charged, and in case 3 positively
charged.  Three different fake matrices were constructed to account
for these three cases, one in which both charges are used, and one
with only negatively or positively charged TLs. When a TL and an
oppositely-charged ID lepton had an invariant mass consistent with
arising from a $Z$ boson, the same-flavour requirement was
automatically satisfied because the TL is taken to have the same flavour
as the ID lepton. The parent sample with two ID leptons contains all
sources of backgrounds that can enter the signal region with a fake
TL, including those with one or two fake ID leptons.  Thus the
procedure predicts the full background contribution with a fake TL.
A small contribution (2\% of the total), evaluated from the MC
simulation, was included to account for events with a `real' TL and
a fake ID lepton.

A summary of expected backgrounds and selected data events in both
the 3ID lepton and 2ID+TL samples is shown in
Table~\ref{tab:sel}.

\begin{table}[htb]
\begin{center}

  \caption{Expected number of background events, number of selected data events and signal
  efficiency (normalized to all decays of the $W$ and $Z$ bosons), after
  the final event selection. The $\ttbar$ backgrounds correspond to SM decays of the top quarks.
  The third entry in the 2ID+TL column corresponds to the fake TL background and includes all sources of
  events in the left-hand column except $ZZ$, $WZ$, $\ttbar W$ and $\ttbar Z$.} \label{tab:sel}
\vspace*{1em}
\begin{tabular}{|l | r l l | r l l|}
\hline & \multicolumn{3}{|c|}{3ID} & \multicolumn{3}{|c|}{2ID+TL} \\
\hline
$ZZ$ and $WZ$         & 9.5  & $\pm$  & $4.4$  &  1.0  &  $\pm$  &          $^{0.5}_{0.6}$\\
$\ttbar W$ and $\ttbar Z$ & 0.51 & $\pm$ & 0.14 & 0.25 & $\pm$ &  0.05\\
$\ttbar$, $WW$    & 0.07  & $\pm$  & 0.02  & \multirow{4}{*}{ 7.6} & \multirow{4}{*}{$\pm$} & \multirow{4}{*}{2.2} \tabularnewline
$Z+$jets              & 1.7  & $\pm$  & 0.7  & &&\\
Single top            & 0.01 & $\pm$  & 0.01 & &&\\
2+3 fake leptons      & 0.0  & $\pm$ & $^{0.2}_{0.0}$ & && \\
\hline
Expected background   & 11.8 &  $\pm$  & 4.4 &                  8.9 &                  $\pm$ &                  2.3 \\
\hline
Data                  & 8\phantom{.0} &       &                & 8\phantom{.0}    &   & \\
\hline
Signal efficiency     & (0.205 &  $\pm$  & 0.024)\%            & (0.045 & $\pm$ & 0.007)\%\\
\hline
\end{tabular}
\end{center}
\end{table}

\begin{figure}
\begin{center}

\includegraphics[width=0.45\textwidth]{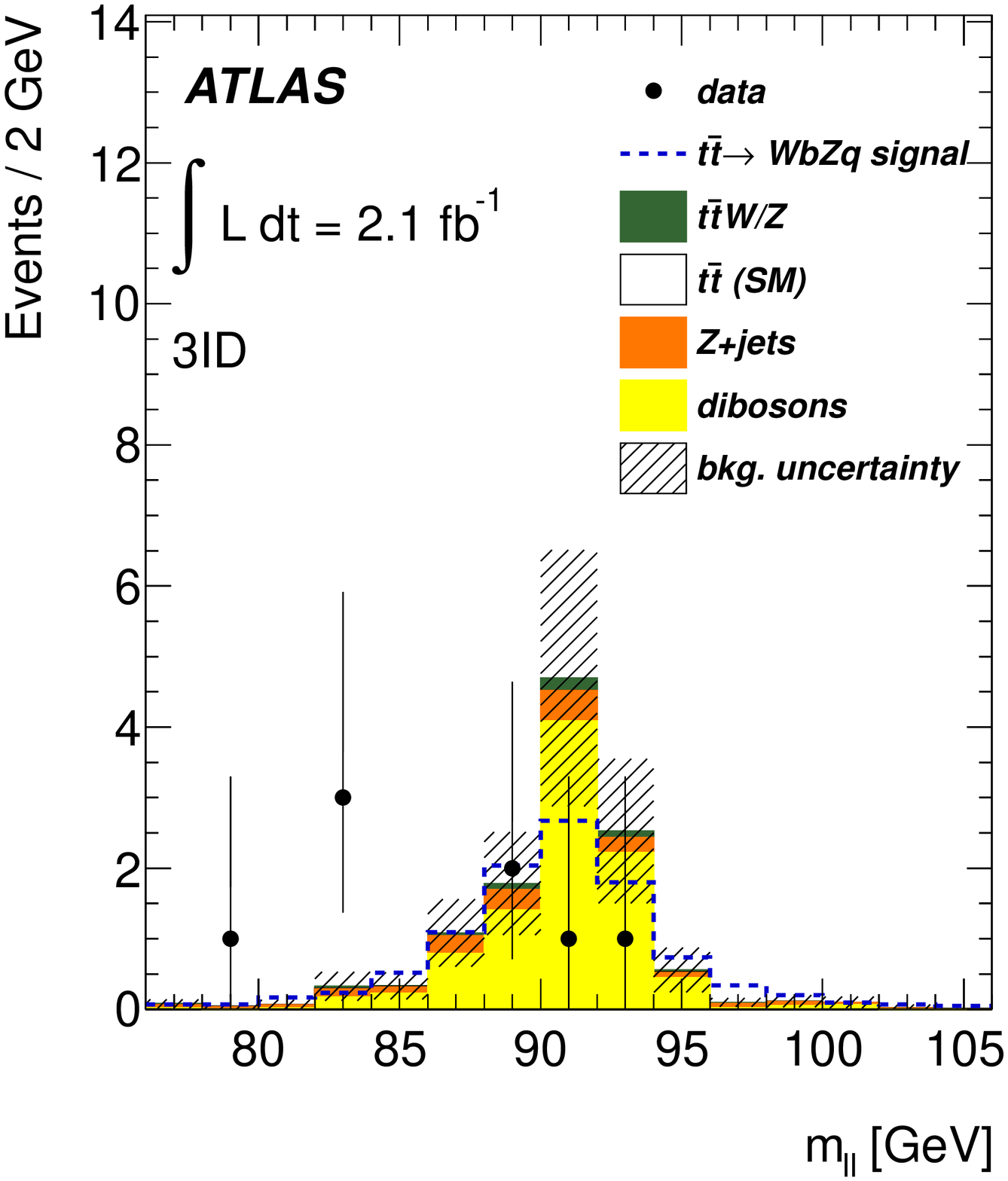}
\includegraphics[width=0.45\textwidth]{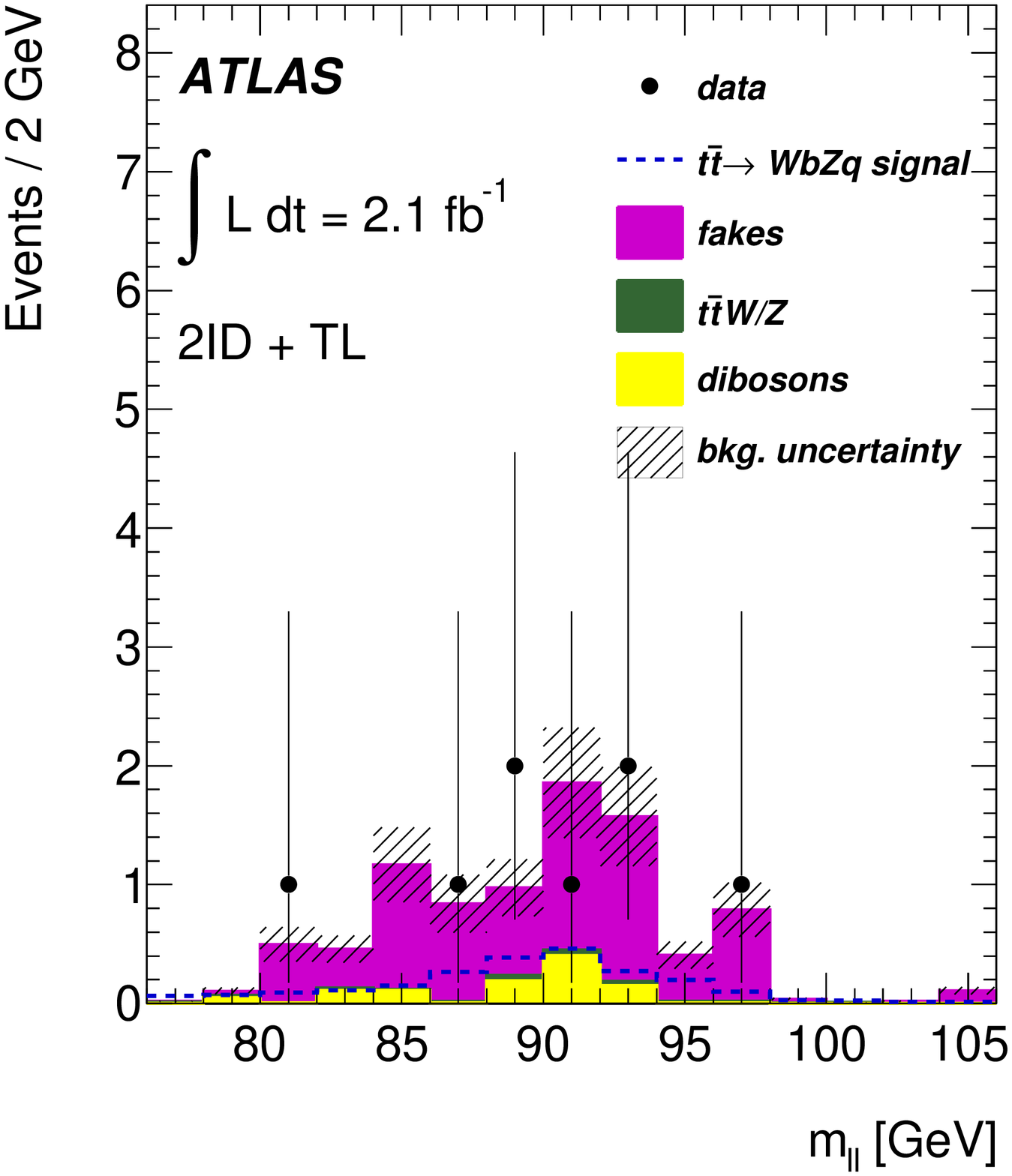}\\
 (a)\hspace*{0.2\textwidth} (b)\hspace*{0.2\textwidth}\\
\includegraphics[width=0.45\textwidth]{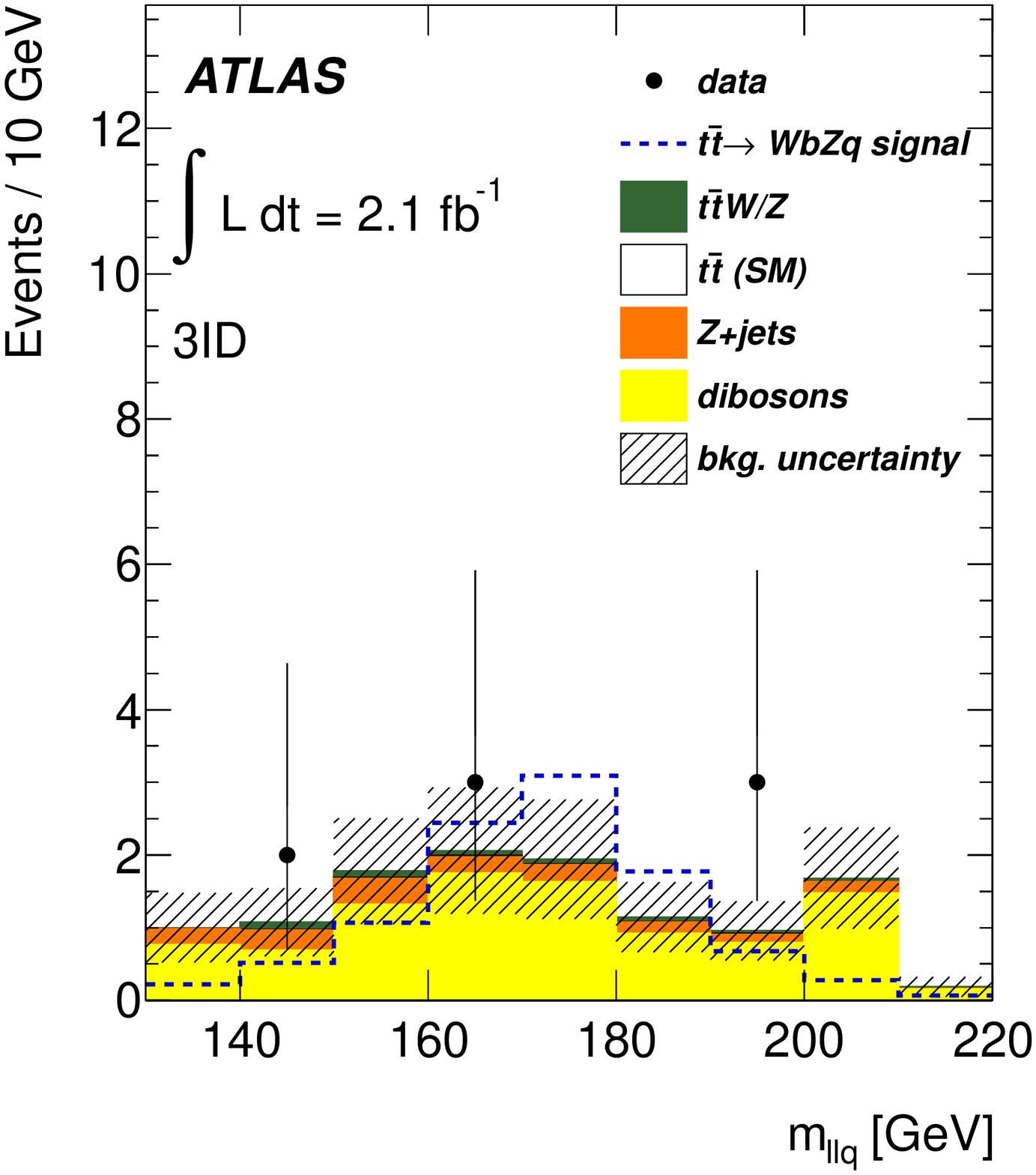}
\includegraphics[width=0.45\textwidth]{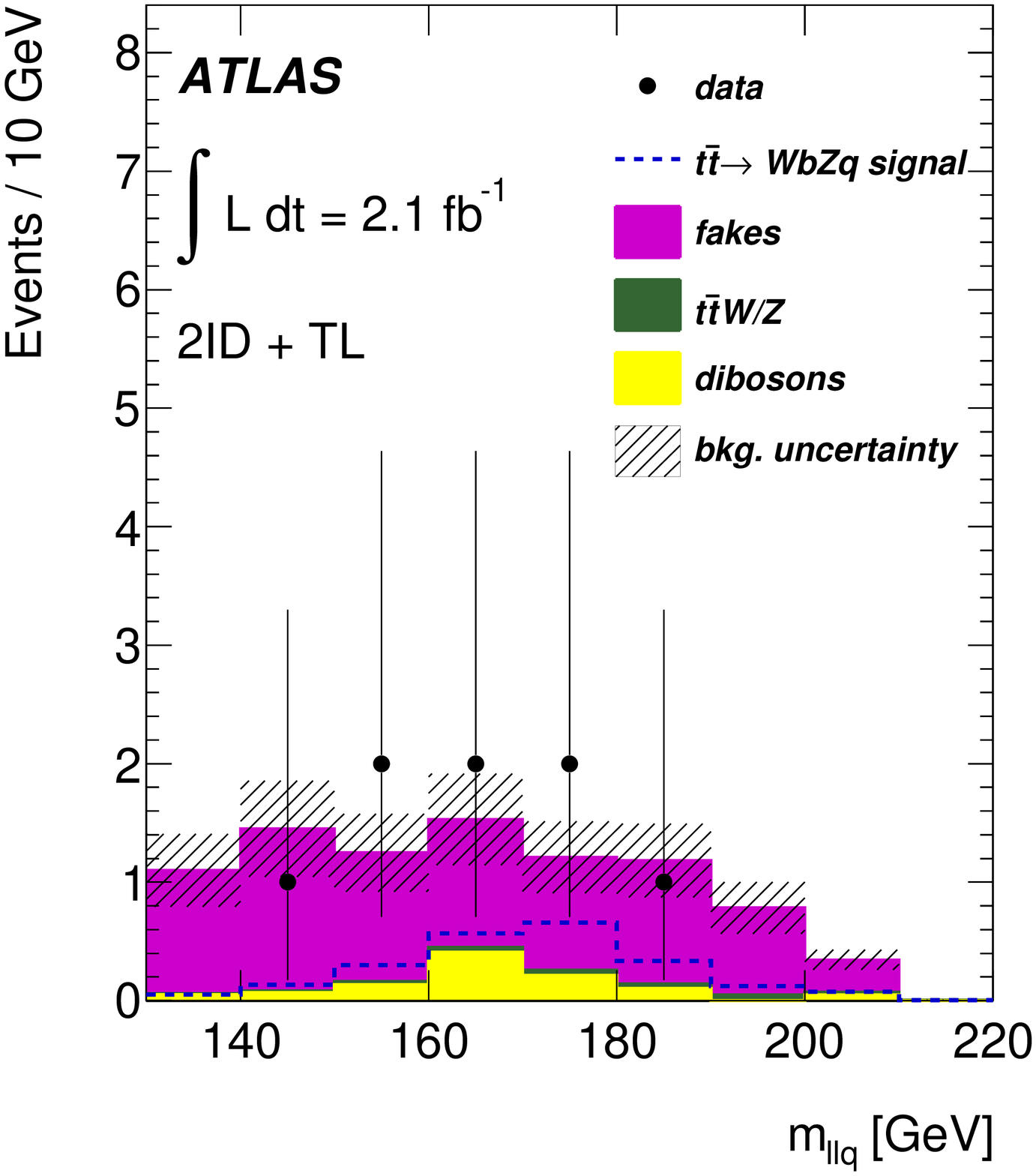}\\

 (c)\hspace*{0.2\textwidth} (d)

\caption{\label{fig:kinsum}Expected and observed $Z$-boson and top-quark
mass distributions for the FCNC decay hypothesis  in the 3ID ((a) \&
(c)) and 2ID+TL ((b) \& (d)) candidate events after all selection
requirements. The $\ttbar\rightarrow WbZq$ distributions are
normalized to the observed limit in each channel.}
\end{center}
\end{figure}

Figure~\ref{fig:kinsum} shows the reconstructed candidate $Z$-boson and top-quark
masses, $m_{ll}$ and $m_{llq}$ respectively, for the FCNC decay hypothesis in the selected candidate events, for both
the 3ID and 2ID+TL data, compared with the expectations from SM
backgrounds and the FCNC signal.

%% file: systematics.tex
\section{Systematic uncertainties\label{sec:syst}}
A number of systematic uncertainties can influence the expected
number of signal and/or background events. The effect of each source
of systematic uncertainty was studied by independently varying the
corresponding central value by the estimated uncertainty. For each
variation, the total number of expected background events and the signal
efficiencies were compared with the reference values.

The measurement of the integrated luminosity has a total uncertainty
of 3.7\%~\cite{lumiPub, lumi}. This uncertainty was considered in the
analyses by changing the normalizations of the backgrounds evaluated
from MC simulation. Uncertainties associated with the energy scale
of light-quark jets and $b$-jets were studied as a function of the jet
transverse momentum and pseudorapidity. These uncertainties, including
the effects of pile-up, are in the range 6--10\%~\cite{jetcor}. The
effects of the jet reconstruction efficiency uncertainty were
studied by randomly removing about 2\% of jets from the
events. 
 The effect of potential jet
resolution mis-modelling in the MC simulation was evaluated by
additional smearing of the reconstructed jet energies within the
uncertainties. In each case, the difference with respect to the
nominal simulation was considered as the systematic uncertainty. The
uncertainties due to MC modelling of the lepton trigger,
reconstruction and selection efficiencies, and $b$-tagging
efficiency, were taken into account by re-computing the predicted
event yields and signal acceptance using the corresponding
systematic shift. The momentum of the lepton in simulation was rescaled and smeared to correct for scale and resolution disagreements between simulated and observed data.
 The systematic
uncertainty associated with the modelling of the momentum scale and
resolution was evaluated by shifting the momentum scale and changing the smearing
factors. Changes applied to electrons, muons and jets were
propagated to $\met$. Uncertainties related to $\met$ were also
studied: the effect of the energy in the calorimeter not
associated with the above objects, and of low momentum (7 \gev\
$<\pt <$ 20 \gev\ ) jets, was studied, as well as the uncertainty
due to modelling of pile-up. The effect of a hardware failure in the
electromagnetic calorimeter was also considered as a systematic
uncertainty (LAr readout problem) and evaluated by varying the jet
thresholds used for removing events with jets directed at the dead
region. The effects of ISR/FSR and top-quark mass uncertainties were
evaluated using the MC samples described in
Section~\ref{sec:MC_Samples}. The effect of uncertainties in
the PDF used for signal generation was evaluated by comparing the
signal acceptance using MSTW2008LO with that from MRST2007 LO$^*$
PDFs. The systematic uncertainties related to the $ZZ$ and $WZ$
simulation modelling were estimated using the Berends-Giele
scaling~\cite{Berends:1990ax, Ellis:1985vn} with an uncertainty of
24\% per jet, added in quadrature. An uncertainty of 4\% was
included for the 0-jet bin. The $ZZ$ and $WZ$ cross sections were
varied by their theoretical uncertainty of
5\%~\cite{CampbellEllis2010}. The uncertainties on the $Z$+jets
normalizations were derived using a data-driven method. In the
2ID+TL channel, where $b$-tagging was used, a systematic uncertainty
associated with the heavy-flavour content of $WZ$+jets and $ZZ$+jets
is included.  This was evaluated by comparing {\tt ALPGEN} and {\tt
MC@NLO}, and is small because the dominant source of $b$-tags in
these events comes from mis-tags of light-quark jets, with a
secondary component from charm jets.

 The dominant source of systematic
uncertainty for the 3ID channel is the $ZZ$ and $WZ$ simulation
modelling. The other sources have effects at most of the same magnitude as the statistical uncertainty.
 The
dominant uncertainty in the 2ID+TL channel is the systematic
uncertainty on the fake-TL prediction, because 90\% of the expected
background arises from this source. This was determined to be 20\%
by comparing predicted and observed events with TLs in control
regions dominated by fake TLs~\cite{DILPaper}.

The resulting uncertainties for the backgrounds and signal
acceptance are shown in Table~\ref{tab:systematics}.  Because almost
90\% of the 2ID+TL background evaluation is data-driven, the 2ID+TL analysis has a
smaller relative background systematic uncertainties in most
categories, compared to the 3ID analysis.

\begin{table}[tb]
\begin{center}
  \caption{Relative changes in the expected number of
  background events and signal yield for different sources of
  systematic uncertainties. The contributions from the $ZZ$ and $WZ$ event generator apply only to the
  simulated background samples.}
  \label{tab:systematics}

\vspace*{1em}
\begin{tabular}{|lrr|rr|}
\hline
& \multicolumn{2}{c}{~~3ID} & \multicolumn{2}{c|}{~2ID+TL}\\ \hline
Source                           & Background & Signal & Background & Signal\\
\hline
Luminosity                       &        4\% &          4\% &          $<$1\% &     4\% \\
\hline
Electron trigger                 &        4\% &          1\% &          $<$1\% &  $<$1\% \\
Electron reconstruction modelling &       10\% &          3\% &          $<$1\% &     2\% \\
\hline
Muon trigger                     &        3\% &          1\% &          $<$1\% &  $<$1\% \\
Muon reconstruction modelling     &        7\% &          1\% &          $<$1\% &     1\% \\
TL reconstruction modelling       &        --- &         ---  &             2\% &     1\% \\
\hline
Jet energy scale                 &       11\% &          1\% &             1\% &     1\% \\
Jet reconstruction efficiency             &        5\% &          2\% &          $<$1\% &  $<$1\% \\
Jet energy resolution            &        1\% &          3\% &             1\% &     4\% \\
\hline
$\met$ modelling                  &        4\% &          1\% &          $<$1\% &  $<$1\% \\
LAr readout problem              &        3\% &          1\% &          $<$1\% &     1\% \\
Pile-up                          &        4\% &       $<$1\% &          $<$1\% &  $<$1\% \\
$b$-tagging                        &        --- &          --- &          1\% &     6\% \\
\hline
Top quark mass                   &     $<$1\% &          2\% &           ---~~ &     3\% \\
$\sigma_{t\bar t}$               &     $<$1\% &          8\% &           ---~~ &     8\% \\
ISR/FSR                          &     $<$1\% &          3\% &           ---~~ &     6\% \\
PDFs                             &      ---~~ &          3\% &           ---~~ &     3\% \\
\hline
$ZZ$ and $WZ$ shape              &       33\% &        ---~~ &             5\% &   ---~~ \\
$ZZ$ and $WZ$ cross section      &        4\% &        ---~~ &             $<$1\% &   ---~~ \\
 $ZZ$ and $WZ$ heavy-flavour content &   ---~~    &   ---~~   & $<$1\%       & ---~~\\
 \hline
Fake leptons                     &        1\% &        ---~~ &            17\% &   ---~~ \\
Total                            &       38\% &         12\% &            18\%    &   15\% \\
\hline
\end{tabular}
\end{center}
\end{table}

%% file: limits.tex
\section{Limit evaluation\label{sec:lim}}

Good agreement between data and expected background yields was
observed, as shown in Table~\ref{tab:sel}. No evidence for the $t\to
Zq$ decay mode was found and 95\% C.L. upper limits on the number of signal
events were derived using the modified frequentist (CL$_{\mathrm
s}$) likelihood method~\cite{Junk:1999kv, alexread}.  The
statistical fluctuations of the pseudo-experiments were performed
using Poisson distributions. All statistical and systematic
uncertainties of the expected backgrounds and signal efficiencies
were taken into account, as described in Section~\ref{sec:syst} and
were implemented assuming Gaussian distributions~\cite{Junk:1999kv}.
The systematic uncertainties of the $ZZ$, $WZ$ and signal acceptance
were considered to be fully correlated between the 3ID and 2ID+TL
channels, while all other sources of uncertainties (statistical or
systematic) were considered uncorrelated. The limits on the number
of signal events were converted into upper limits on the
corresponding BRs using the approximate NNLO calculation, and its
uncertainty, for the $\ttbar$ cross section
($\sigma_{\ttbar}=165^{+11}_{-16}$~pb)~\cite{Langenfeld:2009tc}, and
constraining BR$(t\to Wb)=1-$BR$(t\to Zq)$. The observed 95\% C.L.
upper limit on the FCNC $t\rightarrow Zq$ BR is 0.82\% (3.2\%)
taking the 3ID (2ID+TL) events and background evaluation alone, and
0.73\% when the 3ID and 2ID+TL results are combined.
Table~\ref{qz3l:tab:limits} shows the observed and expected limits
in the absence of signal for the 3ID and 2ID+TL channels, as well as
for the combination. Also shown are the $\pm 1\sigma$ expected
limits.

\begin{table}[tb]
\begin{center}
  \caption{The expected and observed 95\% C.L. upper limits on the FCNC top quark decay $t\to
  Zq$ BR. The $\pm 1\sigma$ expected limits include both statistical and systematic uncertainties. 
}
  \label{qz3l:tab:limits}

\vspace*{1em}
\begin{tabular}{lcccccc}
\hline
channel         & observed & $(-1\sigma)$ & expected & $(+1\sigma)$ \\

\hline
3ID         & 0.81\% & 0.63\% & 0.95\% & 1.4\% \\
2ID+TL      & 3.2\% & 2.15\% & 3.31\% & 4.9\% \\
Combination & 0.73\% & 0.61\% & 0.93\% & 1.4\% \\

\hline
\end{tabular}
\end{center}
\end{table}

%% file: conclusions.tex
\section{Conclusions\label{sec:conclusions}}

A search for FCNC decays of top quarks produced in pairs was
performed using data collected by the ATLAS experiment at a
centre-of-mass energy of $\sqrt{s}=7$~TeV and corresponding to an
integrated luminosity of 2.1~fb$^{-1}$. The search for the $t\to qZ$
decay mode was performed by studying top-quark pair production with
one top quark decaying according to the Standard Model and the other
according to the FCNC ($t\bar t\to bWqZ$). No evidence for such a
signal was found. An observed limit at 95\% C.L. on the $t\to qZ$
FCNC top-quark decay branching fraction was set at BR$(t\to
qZ)<0.73$\%, assuming BR$(t\to bW)+$BR$(t\to qZ)=1$. The observed
limit is compatible with the expected sensitivity, assuming that the
data are described correctly by the Standard Model, of BR$(t\to
qZ)<0.93$\%.

%% file: Acknowledgement-09Feb12-collisions.tex
\section{Acknowledgements}

We thank CERN for the very successful operation of the LHC, as well as the
support staff from our institutions without whom ATLAS could not be
operated efficiently.

We acknowledge the support of ANPCyT, Argentina; YerPhI, Armenia; ARC,
Australia; BMWF, Austria; ANAS, Azerbaijan; SSTC, Belarus; CNPq and FAPESP,
Brazil; NSERC, NRC and CFI, Canada; CERN; CONICYT, Chile; CAS, MOST and NSFC,
China; COLCIENCIAS, Colombia; MSMT CR, MPO CR and VSC CR, Czech Republic;
DNRF, DNSRC and Lundbeck Foundation, Denmark; EPLANET and ERC, European Union;
IN2P3-CNRS, CEA-DSM/IRFU, France; GNAS, Georgia; BMBF, DFG, HGF, MPG and AvH
Foundation, Germany; GSRT, Greece; ISF, MINERVA, GIF, DIP and Benoziyo Center,
Israel; INFN, Italy; MEXT and JSPS, Japan; CNRST, Morocco; FOM and NWO,
Netherlands; RCN, Norway; MNiSW, Poland; GRICES and FCT, Portugal; MERYS
(MECTS), Romania; MES of Russia and ROSATOM, Russian Federation; JINR; MSTD,
Serbia; MSSR, Slovakia; ARRS and MVZT, Slovenia; DST/NRF, South Africa;
MICINN, Spain; SRC and Wallenberg Foundation, Sweden; SER, SNSF and Cantons of
Bern and Geneva, Switzerland; NSC, Taiwan; TAEK, Turkey; STFC, the Royal
Society and Leverhulme Trust, United Kingdom; DOE and NSF, United States of
America.

The crucial computing support from all WLCG partners is acknowledged
gratefully, in particular from CERN and the ATLAS Tier-1 facilities at
TRIUMF (Canada), NDGF (Denmark, Norway, Sweden), CC-IN2P3 (France),
KIT/GridKA (Germany), INFN-CNAF (Italy), NL-T1 (Netherlands), PIC (Spain),
ASGC (Taiwan), RAL (UK) and BNL (USA) and in the Tier-2 facilities
worldwide.

%% file: atlas_authlist.tex
\begin{flushleft}
{\Large The ATLAS Collaboration}

\bigskip

G.~Aad$^{\rm 48}$,
B.~Abbott$^{\rm 111}$,
J.~Abdallah$^{\rm 11}$,
S.~Abdel~Khalek$^{\rm 115}$,
A.A.~Abdelalim$^{\rm 49}$,
O.~Abdinov$^{\rm 10}$,
B.~Abi$^{\rm 112}$,
M.~Abolins$^{\rm 88}$,
O.S.~AbouZeid$^{\rm 158}$,
H.~Abramowicz$^{\rm 153}$
% ...
\bigskip

% ...
$^{10}$ Institute of Physics, Azerbaijan Academy of Sciences, Baku, Azerbaijan\\
$^{11}$ Institut de F\'isica d'Altes Energies and Departament de F\'isica de la Universitat Aut\`onoma  de Barcelona and ICREA, Barcelona, Spain\\
% ...
$^{48}$ Fakult\"{a}t f\"{u}r Mathematik und Physik, Albert-Ludwigs-Universit\"{a}t, Freiburg i.Br., Germany\\
$^{49}$ Section de Physique, Universit\'e de Gen\`eve, Geneva, Switzerland\\
% ...
$^{88}$ Department of Physics and Astronomy, Michigan State University, East Lansing MI, United States of America\\
% ...
$^{111}$ Homer L. Dodge Department of Physics and Astronomy, University of Oklahoma, Norman OK, United States of America\\
$^{112}$ Department of Physics, Oklahoma State University, Stillwater OK, United States of America\\
% ...
$^{115}$ LAL, Universit\'e Paris-Sud and CNRS/IN2P3, Orsay, France\\
% ...
$^{153}$ Raymond and Beverly Sackler School of Physics and Astronomy, Tel Aviv University, Tel Aviv, Israel\\
% ...
$^{158}$ Department of Physics, University of Toronto, Toronto ON, Canada\\
% ...
\end{flushleft}


%% file: main.bbl
\providecommand{\href}[2]{#2}\begingroup\raggedright\begin{thebibliography}{10}

\bibitem{TevEWKWG:2011wr}
{\bf CDF and D0} Collaboration, {\it {Combination of CDF and D0 results on the
  mass of the top quark using up to 5.8 fb$^{-1}$ of data}},
  \href{http://xxx.lanl.gov/abs/1107.5255}{{\tt arXiv:1107.5255}}.

\bibitem{Glashow:1970gm}
S.~L. Glashow, J.~Iliopoulos, and L.~Maiani, {\it {Weak Interactions with
  Lepton-Hadron Symmetry}},  {\em \rm Phys. Rev.} {\bf D2} (1970) 1285.

\bibitem{AguilarSaavedra:2002ns}
J.~A. Aguilar-Saavedra and B.~M. Nobre, {\it {Rare top decays $t\to c \gamma,
  t\to c g$ and CKM unitarity}},  {\em \rm Phys. Lett.} {\bf B553} (2003) 251,
  [\href{http://xxx.lanl.gov/abs/hep-ph/0210360}{{\tt hep-ph/0210360}}].

\bibitem{delAguila:1998tp}
F.~del Aguila, J.~A. Aguilar-Saavedra, and R.~Miquel, {\it {Constraints on top
  couplings in models with exotic quarks}},  {\em \rm Phys. Rev. Lett.} {\bf
  82} (1999) 1628, [\href{http://xxx.lanl.gov/abs/hep-ph/9808400}{{\tt
  hep-ph/9808400}}].

\bibitem{AguilarSaavedra:2002kr}
J.~A. Aguilar-Saavedra, {\it {Effects of mixing with quark singlets}},  {\em
  \rm Phys. Rev.} {\bf D67} (2003) 035003,
  [\href{http://xxx.lanl.gov/abs/hep-ph/0210112}{{\tt hep-ph/0210112}}].
  Erratum-ibid. {\bf D69} (2004) 099901.

\bibitem{cheng:1987rs}
T.~P. Cheng and M.~Sher, {\it {Mass Matrix Ansatz and Flavor Nonconservation in
  Models with Multiple Higgs Doublets}},  {\em \rm Phys. Rev.} {\bf D35} (1987)
  3484.

\bibitem{Grzadkowski:1990sm}
B.~Grzadkowski, J.~F. Gunion, and P.~Krawczyk, {\it {Neutral current flavor
  changing decays for the Z boson and the top quark in two Higgs doublet
  models}},  {\em \rm Phys. Lett.} {\bf B268} (1991) 106.

\bibitem{Luke:1993cy}
M.~E. Luke and M.~J. Savage, {\it {Flavor changing neutral currents in the
  Higgs sector and rare top decays}},  {\em \rm Phys. Lett.} {\bf B307} (1993)
  387, [\href{http://xxx.lanl.gov/abs/hep-ph/9303249}{{\tt hep-ph/9303249}}].

\bibitem{Atwood:1995ud}
D.~Atwood, L.~Reina, and A.~Soni, {\it {Probing flavor changing top - charm -
  scalar interactions in $e^+ e^-$ collisions}},  {\em \rm Phys. Rev.} {\bf
  D53} (1996) 1199, [\href{http://xxx.lanl.gov/abs/hep-ph/9506243}{{\tt
  hep-ph/9506243}}].

\bibitem{Atwood:1996vj}
D.~Atwood, L.~Reina, and A.~Soni, {\it {Phenomenology of two Higgs doublet
  models with flavor changing neutral currents}},  {\em \rm Phys. Rev.} {\bf
  D55} (1997) 3156, [\href{http://xxx.lanl.gov/abs/hep-ph/9609279}{{\tt
  hep-ph/9609279}}].

\bibitem{Bejar:2000ub}
S.~Bejar, J.~Guasch, and J.~Sola, {\it {Loop induced flavor changing neutral
  decays of the top quark in a general two-Higgs-doublet model}},  {\em \rm
  Nucl. Phys.} {\bf B600} (2001) 21,
  [\href{http://xxx.lanl.gov/abs/hep-ph/0011091}{{\tt hep-ph/0011091}}].

\bibitem{Li:1993mg}
C.~S. Li, R.~J. Oakes, and J.~M. Yang, {\it {Rare decay of the top quark in the
  minimal supersymmetric model}},  {\em \rm Phys. Rev.} {\bf D49} (1994) 293.
  Erratum-ibid.D56:3156,1997.

\bibitem{deDivitiis:1997sh}
G.~M. de~Divitiis, R.~Petronzio, and L.~Silvestrini, {\it {Flavour changing top
  decays in supersymmetric extensions of the standard model}},  {\em \rm Nucl.
  Phys.} {\bf B504} (1997) 45,
  [\href{http://xxx.lanl.gov/abs/hep-ph/9704244}{{\tt hep-ph/9704244}}].

\bibitem{Lopez:1997xv}
J.~L. Lopez, D.~V. Nanopoulos, and R.~Rangarajan, {\it {New supersymmetric
  contributions to $t\to c V$}},  {\em \rm Phys. Rev.} {\bf D56} (1997) 3100,
  [\href{http://xxx.lanl.gov/abs/hep-ph/9702350}{{\tt hep-ph/9702350}}].

\bibitem{Guasch:1999jp}
J.~Guasch and J.~Sola, {\it {FCNC top quark decays: A door to SUSY physics in
  high luminosity colliders?}},  {\em \rm Nucl. Phys.} {\bf B562} (1999) 3,
  [\href{http://xxx.lanl.gov/abs/hep-ph/9906268}{{\tt hep-ph/9906268}}].

\bibitem{Delepine:2004hr}
D.~Delepine and S.~Khalil, {\it {Top flavour violating decays in general
  supersymmetric models}},  {\em \rm Phys. Lett.} {\bf B599} (2004) 62,
  [\href{http://xxx.lanl.gov/abs/hep-ph/0406264}{{\tt hep-ph/0406264}}].

\bibitem{Liu:2004qw}
J.~J. Liu, C.~S. Li, L.~L. Yang, and L.~G. Jin, {\it {$t\to c V$ via SUSY FCNC
  couplings in the unconstrained MSSM}},  {\em \rm Phys. Lett.} {\bf B599}
  (2004) 92, [\href{http://xxx.lanl.gov/abs/hep-ph/0406155}{{\tt
  hep-ph/0406155}}].

\bibitem{Cao:2007dk}
J.~J. Cao et~al., {\it {SUSY-induced FCNC top-quark processes at the Large
  Hadron Collider}},  {\em \rm Phys. Rev.} {\bf D75} (2007) 075021,
  [\href{http://xxx.lanl.gov/abs/hep-ph/0702264}{{\tt hep-ph/0702264}}].

\bibitem{Yang:1997dk}
J.~M. Yang, B.-L. Young, and X.~Zhang, {\it {Flavor-changing top quark decays
  in R-parity violating SUSY}},  {\em \rm Phys. Rev.} {\bf D58} (1998) 055001,
  [\href{http://xxx.lanl.gov/abs/hep-ph/9705341}{{\tt hep-ph/9705341}}].

\bibitem{Lu:2003yr}
G.~Lu, F.~Yin, X.~Wang, and L.~Wan, {\it {The rare top quark decays $t \to c V$
  in the topcolor- assisted technicolor model}},  {\em \rm Phys. Rev.} {\bf
  D68} (2003) 015002, [\href{http://xxx.lanl.gov/abs/hep-ph/0303122}{{\tt
  hep-ph/0303122}}].

\bibitem{Agashe:2004cp}
G.~P. K.~Agashe and A.~Soni, {\it {Flavor structure of warped extra dimension
  models}},  {\em \rm Phys. Rev.} {\bf D71} (2005) 016002,
  [\href{http://xxx.lanl.gov/abs/hep-ph/0408134}{{\tt hep-ph/0408134}}].

\bibitem{Agashe:2006wa}
G.~P. K.~Agashe and A.~Soni, {\it {Collider Signals of Top Quark Flavor
  Violation from a Warped Extra Dimension}},  {\em \rm Phys. Rev. D} {\bf 75}
  (2007) 015002, [\href{http://xxx.lanl.gov/abs/hep-ph/0606293}{{\tt
  hep-ph/0606293}}].

\bibitem{Abe:1997fz}
{\bf CDF} Collaboration, F.~Abe et~al., {\it {Search for flavor-changing
  neutral current decays of the top quark in $p \bar{p}$ collisions at
  $\sqrt{s} = 1.8$ TeV}},  {\em \rm Phys. Rev. Lett.} {\bf 80} (1998) 2525.

\bibitem{Abazov:2011qf}
{\bf D0} Collaboration, V.~M. Abazov et~al., {\it {Search for flavor changing
  neutral currents in decays of top quarks}},  {\em \rm Phys. Lett. B} {\bf
  701} (2011) 313.

\bibitem{Heister:2002xv}
{\bf ALEPH} Collaboration, A.~Heister et~al., {\it {Search for single top
  production in $e^+e^-$ collisions at $\sqrt{s}$ up to 209 \gev}},  {\em \rm
  Phys. Lett.} {\bf B543} (2002) 173,
  [\href{http://xxx.lanl.gov/abs/hep-ex/0206070}{{\tt hep-ex/0206070}}].

\bibitem{Abdallah:2003wf}
{\bf DELPHI} Collaboration, J.~Abdallah et~al., {\it {Search for single top
  production via FCNC at LEP at $\sqrt{s}$ = 189 \gev\ - 208 \gev}},  {\em \rm
  Phys. Lett.} {\bf B590} (2004) 21,
  [\href{http://xxx.lanl.gov/abs/hep-ex/0404014}{{\tt hep-ex/0404014}}].

\bibitem{Abbiendi:2001wk}
{\bf OPAL} Collaboration, G.~Abbiendi et~al., {\it {Search for single top quark
  production at LEP2}},  {\em \rm Phys. Lett.} {\bf B521} (2001) 181,
  [\href{http://xxx.lanl.gov/abs/hep-ex/0110009}{{\tt hep-ex/0110009}}].

\bibitem{Achard:2002vv}
{\bf L3} Collaboration, P.~Achard et~al., {\it {Search for single top
  production at LEP}},  {\em \rm Phys. Lett.} {\bf B549} (2002) 290,
  [\href{http://xxx.lanl.gov/abs/hep-ex/0210041}{{\tt hep-ex/0210041}}].

\bibitem{LEPExoticaWG200101}
{The LEP Exotica WG}, {\it {Search for single top production via flavour
  changing neutral currents: preliminary combined results of the LEP
  experiments}},  \href{http://xxx.lanl.gov/abs/LEP Exotica WG 2001-01}{{\tt
  LEP Exotica WG 2001-01}}.

\bibitem{Beneke:2000hk}
M.~Beneke et~al., {\it {Top quark physics, in: Proceedings of the 1999 CERN
  Workshop on SM physics (and more) at the LHC}},
  \href{http://xxx.lanl.gov/abs/hep-ph/0003033}{{\tt hep-ph/0003033}}.

\bibitem{Abramowicz:2011tv}
{\bf ZEUS} Collaboration, H.~Abramowicz et~al., {\it {Search for single-top
  production in $ep$ collisions at HERA}},  {\em \rm Phys.Lett.} {\bf B708}
  (2012) 27, [\href{http://xxx.lanl.gov/abs/1111.3901}{{\tt arXiv:1111.3901}}].

\bibitem{Aktas:2003yd}
{\bf H1} Collaboration, A.~Aktas et~al., {\it {Search for single top quark
  production in $ep$ collisions at HERA}},  {\em \rm Eur. Phys. J.} {\bf C33}
  (2004) 9, [\href{http://xxx.lanl.gov/abs/hep-ex/0310032}{{\tt
  hep-ex/0310032}}].

\bibitem{Aad2012351}
{\bf ATLAS} Collaboration, {\it Search for {FCNC} single top-quark production
  at with the {ATLAS} detector},  {\em \rm Phys. Lett.} {\bf B712} (2012) 351,
  [\href{http://xxx.lanl.gov/abs/1203.0529}{{\tt arXiv:1203.0529}}].

\bibitem{Aad:2008zzm}
{\bf ATLAS} Collaboration, {\it {The ATLAS Experiment at the CERN Large Hadron
  Collider}},  {\em \rm JINST} {\bf 3} (2008) S08003.

\bibitem{lumiPub}
{\bf ATLAS} Collaboration, {\it {Luminosity Determination in $pp$ Collisions at
  $\sqrt{s}$=7 TeV using the ATLAS Detector at the LHC}},  {\em \rm Eur. Phys.
  J.} {\bf C71} (2011) 1630, [\href{http://xxx.lanl.gov/abs/1101.2185}{{\tt
  arXiv:1101.2185}}].

\bibitem{lumi}
{\bf ATLAS} Collaboration, {\it {Luminosity Determination in $pp$ Collisions at
  $\sqrt{s}$=7 TeV using the ATLAS Detector in 2011}},  {\em \rm
  ATLAS-CONF-2011-116}
  [\href{http://xxx.lanl.gov/abs/http://cdsweb.cern.ch/record/1376384}{{\tt
  http://cdsweb.cern.ch/record/1376384}}].

\bibitem{Agostinelli2003250}
S.~Agostinelli et~al., {\it {GEANT4}-a simulation toolkit},  {\em \rm Nucl.
  Instr. Meth.} {\bf A506} (2003) 250.

\bibitem{ATLAS_SIM}
{\bf ATLAS} Collaboration, {\it The {ATLAS} simulation infrastructure},  {\em
  \rm Eur. Phys. J.} {\bf C70} (2010) 823,
  [\href{http://xxx.lanl.gov/abs/1005.4568}{{\tt arXiv:1005.4568}}].

\bibitem{Slabospitsky:2002ag}
S.~R. Slabospitsky and L.~Sonnenschein, {\it {TopReX generator (version 3.25):
  Short manual}},  {\em Comput. Phys. Commun.} {\bf 148} (2002) 87--102,
  [\href{http://xxx.lanl.gov/abs/hep-ph/0201292}{{\tt hep-ph/0201292}}].

\bibitem{Sherstnev:2007nd}
A.~Sherstnev and R.~Thorne, {\it {Parton Distributions for LO Generators}},
  {\em \rm Eur. Phys. J.} {\bf C55} (2008) 553,
  [\href{http://xxx.lanl.gov/abs/0711.2473}{{\tt arXiv:0711.2473}}].

\bibitem{Sjostrand:2006za}
T.~Sjostrand, S.~Mrenna, and P.~Skands, {\it {PYTHIA 6.4 physics and manual}},
  {\em \rm JHEP} {\bf 05} (2006) 026,
  [\href{http://xxx.lanl.gov/abs/hep-ph/0603175}{{\tt hep-ph/0603175}}].

\bibitem{Kersevan:2004yg}
B.~P. Kersevan and E.~Richter-Was, {\it {The Monte Carlo event generator AcerMC
  version 2.0 with interfaces to PYTHIA 6.2 and HERWIG 6.5}},
  \href{http://xxx.lanl.gov/abs/hep-ph/0405247}{{\tt hep-ph/0405247}}.

\bibitem{Skands}
P.~Skands, {\it {Tuning Monte Carlo Generators: The Perugia Tunes}},  {\em \rm
  Phys. Rev.} {\bf D82} (2010) 074018.

\bibitem{alpgenRepeated}
M.~L. Mangano, M.~Moretti, F.~Piccinini, R.~Pittau, and A.~D. Polosa, {\it
  {ALPGEN, a generator for hard multiparton processes in hadronic collisions}},
   {\em \rm JHEP} {\bf 07} (2003) 001.

\bibitem{cteq6l}
J.~Pumplin et~al., {\it {New generation of parton distributions with
  uncertainties from global QCD analysis}},  {\em \rm JHEP} {\bf 07} (2002)
  012.

\bibitem{herwig1}
G.~Corcella et~al., {\it {HERWIG 6.5: an event generator for Hadron Emission
  Reactions With Interfering Gluons (including supersymmetric processes)}},
  {\em JHEP} {\bf 01} (2001) 010.

\bibitem{Corcella:2002jc}
G.~Corcella et~al., {\it {HERWIG 6.5 release note}},
  \href{http://xxx.lanl.gov/abs/hep-ph/0210213}{{\tt hep-ph/0210213}}.

\bibitem{Butterworth:1996zw}
J.~M. Butterworth, J.~R. Forshaw, and M.~H. Seymour, {\it {Multiparton
  interactions in photoproduction at HERA}},  {\em \rm Z. Phys.} {\bf C72}
  (1996) 637, [\href{http://xxx.lanl.gov/abs/hep-ph/9601371}{{\tt
  hep-ph/9601371}}].

\bibitem{PUB-2010-014}
{\bf ATLAS} Collaboration, {\it {First tuning of HERWIG/JIMMY to ATLAS data}},
  \href{http://xxx.lanl.gov/abs/ATLAS-PHYS-PUB-2010-014}{{\tt
  ATLAS-PHYS-PUB-2010-014}}.

\bibitem{mcatnlo1}
S.~Frixione and B.~R. Webber, {\it {Matching NLO QCD computations and parton
  shower simulations}},  {\em \rm JHEP} {\bf 06} (2002) 029,
  [\href{http://xxx.lanl.gov/abs/hep-ph/0204244}{{\tt hep-ph/0204244}}].

\bibitem{Frixione:2003ei}
S.~Frixione, P.~Nason, and B.~R. Webber, {\it {Matching NLO QCD and parton
  showers in heavy flavour production}},  {\em \rm JHEP} {\bf 08} (2003) 007,
  [\href{http://xxx.lanl.gov/abs/hep-ph/0305252}{{\tt hep-ph/0305252}}].

\bibitem{Frixione:2005vw}
S.~Frixione, E.~Laenen, P.~Motylinski, and B.~R. Webber, {\it {Single-top
  production in MC@NLO}},  {\em \rm JHEP} {\bf 03} (2006) 092,
  [\href{http://xxx.lanl.gov/abs/hep-ph/0512250}{{\tt hep-ph/0512250}}].

\bibitem{Nadolsky:2008zw}
P.~M. Nadolsky et~al., {\it {Implications of CTEQ global analysis for collider
  observables}},  {\em \rm Phys.Rev.} {\bf D78} (2008) 013004,
  [\href{http://xxx.lanl.gov/abs/0802.0007}{{\tt arXiv:0802.0007}}].

\bibitem{cteq66}
P.~M. Nadolsky et~al., {\it {Implications of CTEQ global analysis for collider
  observables}},  {\em \rm Phys. Rev.} {\bf D78} (2008) 013004,
  [\href{http://xxx.lanl.gov/abs/0802.0007}{{\tt arXiv:0802.0007}}].

\bibitem{Aliev:2010}
M.~Aliev et~al., {\it {HATHOR -- HAdronic Top and Heavy quarks crOss section
  calculatoR}},  {\em \rm Comput. Phys. Commun.} {\bf 182} (2011) 1034.

\bibitem{Kidonakis:2011wy}
N.~Kidonakis, {\it {Next-to-next-to-leading-order collinear and soft gluon
  corrections for t-channel single top quark production}},  {\em \rm Phys.
  Rev.} {\bf D83} (2011) 091503, [\href{http://xxx.lanl.gov/abs/1103.2792}{{\tt
  arXiv:1103.2792}}].

\bibitem{Kidonakis:2010tc}
N.~Kidonakis, {\it {NNLL resummation for s-channel single top quark
  production}},  {\em \rm Phys. Rev.} {\bf D81} (2010) 054028.

\bibitem{Kidonakis:2010ux}
N.~Kidonakis, {\it {Two-loop soft anomalous dimensions for single top quark
  associated production with a W- or H-}},  {\em \rm Phys. Rev.} {\bf D82}
  (2010) 054018.

\bibitem{Alwall:2007st}
J.~Alwall et~al., {\it {MadGraph/MadEvent v4: The New Web Generation}},  {\em
  \rm JHEP} {\bf 0709} (2007) 028,
  [\href{http://xxx.lanl.gov/abs/0706.2334}{{\tt arXiv:0706.2334}}].

\bibitem{DILPaper}
{\bf ATLAS} Collaboration, {\it {Measurement of the cross section for top-quark
  pair production in pp collisions at $\sqrt{s}=$7 TeV with the ATLAS detector
  using final states with two high-$\pt$ leptons}},  {\em \rm JHEP} {\bf 1205}
  (2012) 059, [\href{http://xxx.lanl.gov/abs/1202.4892}{{\tt
  arXiv:1202.4892}}].

\bibitem{ElectronPerformance}
{\bf ATLAS} Collaboration, {\it {Electron performance measurements with the
  ATLAS detector using the 2010 LHC proton-proton collision data}},  {\em \rm
  Eur. Phys. J.} {\bf C72} (2012) 1909,
  [\href{http://xxx.lanl.gov/abs/1110.3174}{{\tt arXiv:1110.3174}}].

\bibitem{antikt}
M.~Cacciari, G.~P. Salam, and G.~Soyez, {\it {The anti-$k_t$\ jet clustering
  algorithm}},  {\em \rm JHEP} {\bf 0804} (2008) 063.

\bibitem{jetcor}
{\bf ATLAS} Collaboration, {\it {Jet energy measurement with the ATLAS detector
  in proton-proton collisions at $\sqrt{s}$ = 7 TeV}},  {\em \rm Submitted to
  Eur. Phys. J.} [\href{http://xxx.lanl.gov/abs/1112.6426}{{\tt
  arXiv:1112.6426}}].

\bibitem{AdvBtag}
{\bf ATLAS} Collaboration, {\it {Commissioning of the ATLAS high-performance
  b-tagging algorithms in the 7 TeV collision data}},  {\em \rm
  ATLAS-CONF-2011-102}
  [\href{http://xxx.lanl.gov/abs/http://cdsweb.cern.ch/record/1369219}{{\tt
  http://cdsweb.cern.ch/record/1369219}}].

\bibitem{METPaper}
{\bf ATLAS} Collaboration, {\it {Performance of missing transverse momentum
  reconstruction in proton-proton collisions at 7 TeV with ATLAS}},  {\em \rm
  Eur. Phys. J.} {\bf C72} (2012) 1844,
  [\href{http://xxx.lanl.gov/abs/1108.5602}{{\tt arXiv:1108.5602}}].

\bibitem{top2010}
{\bf ATLAS} Collaboration, {\it {Measurement of the top quark-pair production
  cross-section with ATLAS in pp collisions at $\sqrt{s} = 7 \TeV$}},  {\em \rm
  Eur. Phys. J.} {\bf C71} (2011) 1577,
  [\href{http://xxx.lanl.gov/abs/1012.1792}{{\tt arXiv:1012.1792}}].

\bibitem{Feldman:1997qc}
G.~J. Feldman and R.~Cousins, {\it {A unified approach to the classical
  statistical analysis of small signals}},  {\em \rm Phys. Rev.} {\bf D57}
  (1998) 3873, [\href{http://xxx.lanl.gov/abs/physics/9711021}{{\tt
  physics/9711021}}].

\bibitem{Berends:1990ax}
F.~A. Berends, H.~Kuijf, B.~Tausk, and W.~T. Giele, {\it {On the Production of
  a W and Jets at Hadron Colliders}},  {\em \rm Nucl. Phys.} {\bf B357} (1991)
  32.

\bibitem{Ellis:1985vn}
S.~D. Ellis, R.~Kleiss, and W.~J. Stirling, {\it {W's, Z's and Jets}},  {\em
  \rm Phys. Lett.} {\bf B154} (1985) 435.

\bibitem{CampbellEllis2010}
J.~M. Campbell and R.~K. Ellis, {\it {MCFM for the Tevatron and the LHC}},
  {\em \rm Nucl. Phys. Proc. Suppl.} {\bf 205} (2010) 10.

\bibitem{Junk:1999kv}
T.~Junk, {\it {Confidence Level Computation for Combining Searches with Small
  Statistics}},  {\em \rm Nucl. Instrum. Meth.} {\bf A434} (1999) 435,
  [\href{http://xxx.lanl.gov/abs/hep-ex/9902006}{{\tt hep-ex/9902006}}].

\bibitem{alexread}
A.~L. Read, {\it {Modified frequentist analysis of search results (The {CL(s)}
  method)}},  \href{http://xxx.lanl.gov/abs/CERN-OPEN-2000-205}{{\tt
  CERN-OPEN-2000-205}}. Prepared for Workshop on Confidence Limits, Geneva,
  Switzerland, 17-18 Jan 2000.

\bibitem{Langenfeld:2009tc}
U.~Langenfeld, S.~Moch, and P.~Uwer, {\it {New Results for \ttbar\ Production
  at Hadron Colliders}},  in {\em XVII International Workshop on Deep-Inelastic
  Scattering and Related Topics, Madrid, Spain}, April, 2009.

\end{thebibliography}\endgroup
